\documentclass[11pt,draftclsnofoot,onecolumn,romanappendices]{IEEEtran}

\IEEEoverridecommandlockouts

\usepackage{amsmath,amsfonts,amssymb,amsthm}
\usepackage{mathrsfs}
\usepackage{dsfont}
\usepackage{comment,blkarray}
\usepackage{multirow,bigdelim}
\usepackage{mathrsfs}
\usepackage{cite}
\usepackage[ruled,commentsnumbered, vlined]{algorithm2e}
\usepackage{tikz}
\usetikzlibrary{arrows,automata}
\usepackage[latin1]{inputenc}
\usepackage{verbatim}
\usepackage{graphicx}
\usepackage{subfigure}		
\usepackage{booktabs}
\usepackage{caption}	
\usepackage{color}
\usepackage{colortbl}
\usepackage{xcolor}												

\includecomment{itwfull}
\excludecomment{DiscussionAchievality}
\excludecomment{details}

\excludecomment{journalonly}

\usepackage{url}
\usepackage{subfigure}
\usepackage{multicol}

\usepackage{ifpdf}															%
\ifpdf																		%
	\usepackage{hyperref}
\else																		%
\fi	

\usetikzlibrary{arrows,decorations.pathmorphing,decorations.footprints,fadings,calc,
trees,mindmap,shadows,decorations.text,patterns,positioning,shapes,matrix,fit}

\makeatletter
\newcommand{\rmnum}[1]{\romannumeral #1}
\newcommand{\Rmnum}[1]{\expandafter\@slowromancap\romannumeral #1@}

\newif\if@borderstar
\def\bordermatrix{\@ifnextchar*{%
  \@borderstartrue\@bordermatrix@i}{\@borderstarfalse\@bordermatrix@i*}%
}
\def\@bordermatrix@i*{\@ifnextchar[{\@bordermatrix@ii}{\@bordermatrix@ii[()]}}
\def\@bordermatrix@ii[#1]#2{%
\begingroup
  \m@th\@tempdima8.75\p@\setbox\z@\vbox{%
    \def\cr{\crcr\noalign{\kern 2\p@\global\let\cr\endline }}%
    \ialign {$##$\hfil\kern 2\p@\kern\@tempdima & \thinspace %
    \hfil $##$\hfil && \quad\hfil $##$\hfil\crcr\omit\strut %
    \hfil\crcr\noalign{\kern -\baselineskip}#2\crcr\omit %
    \strut\cr}}%
  \setbox\tw@\vbox{\unvcopy\z@\global\setbox\@ne\lastbox}%
  \setbox\tw@\hbox{\unhbox\@ne\unskip\global\setbox\@ne\lastbox}%
  \setbox\tw@\hbox{%
    $\kern\wd\@ne\kern -\@tempdima\left\@firstoftwo#1%
    \if@borderstar\kern2pt\else\kern -\wd\@ne\fi%
    \global\setbox\@ne\vbox{\box\@ne\if@borderstar\else\kern 2\p@\fi}%
    \vcenter{\if@borderstar\else\kern -\ht\@ne\fi%
    \unvbox\z@\kern -\if@borderstar2\fi\baselineskip}%
    \if@borderstar\kern -2\@tempdima\kern2\p@\else\,\fi\right\@secondoftwo#1 $%
  }\null \;\vbox{\kern\ht\@ne\box\tw@}%
\endgroup
}
\makeatother

\newtheorem{thm}{Theorem}
\newtheorem{lemma}[thm]{Lemma}
\newtheorem{cor}[thm]{Corollary}

\newtheorem{example}{Example}
\newtheorem{defn}{Definition}
\newtheorem{remark}{Remark}
\allowdisplaybreaks

\newcommand{\ein}{\mathrm{In}}
\newcommand{\eout}{\mathrm{Out}}
\newcommand{\tail}{\mathrm{tail}}
\newcommand{\head}{\mathrm{head}}
\newcommand{\mincut}{\mathrm{mincut}}

\newcommand{\mC}{\mathcal{C}}
\newcommand{\mN}{\mathcal{N}}
\newcommand{\mG}{\mathcal{G}}
\newcommand{\mV}{\mathcal{V}}
\newcommand{\mE}{\mathcal{E}}
\newcommand{\mW}{\mathcal{W}}

\newcommand{\mbC}{\mathbf{C}}
\newcommand{\hmbC}{\widehat{\mathbf{C}}}
\newcommand{\hmC}{\widehat{\mathcal{C}}}
\newcommand{\hC}{\widehat{C}}
\newcommand{\htheta}{\widehat{\theta}}
\newcommand{\hvarphi}{\widehat{\varphi}}
\newcommand{\mA}{\mathcal{A}}
\newcommand{\mB}{\mathcal{B}}

\newcommand{\mO}{\mathcal{O}}
\newcommand{\mK}{\mathcal{K}}
\newcommand{\vX}{\mathbf{X}}
\newcommand{\vx}{\textit{\textbf{x}}}
\newcommand{\bx}{{\bf x}}
\newcommand{\vY}{\mathbf{Y}}
\newcommand{\vy}{\textit{\textbf{y}}}
\newcommand{\vM}{\mathbf{M}}
\newcommand{\vm}{\textit{\textbf{m}}}

\newcommand{\vK}{\mathbf{K}}
\newcommand{\vk}{\textit{\textbf{k}}}
\newcommand{\vb}{\vec{b}}
\newcommand{\vg}{\vec{g}}
\newcommand{\vh}{\vec{h}}
\newcommand{\Fq}{\mathbb{F}_q}

\newcommand{\FqL}{\mathbb{F}_{q^L}}

\newcommand{\bfzero}{{\bf 0}}

\newcommand{\hB}{\widehat{B}}

\newcommand{\Rank}{{\mathrm{Rank}}}
\newcommand{\hg}{\widehat{g}}
\newcommand{\hW}{\widehat{W}}
\newcommand{\bzero}{{\vec{0}}}

\newcommand{\mL}{\mathcal{L}}

\tikzstyle{vertex}=[draw,circle,fill=gray!30,minimum size=6pt, inner sep=0pt]

\excludecomment{fullversion}

\begin{document}
\title{Secure Network Function Computation for Linear Functions\,--\,Part \Rmnum{1}: Source Security}

\author{Xuan~Guang,~Yang~Bai,~and~Raymond~W.~Yeung
\thanks{This paper was presented in part at the 2021 and 2022 IEEE International Symposium on Information Theory.}
}
%\thanks{X. Guang and Y. Bai are with School of Mathematical Sciences and LPMC, Nankai University, Tianjin, China (e-mails: xguang@nankai.edu.cn, bbbyang@mail.nankai.edu.cn); R. W. Yeung is with the Institute of Network Coding and Department of Information Engineering, The Chinese University of Hong Kong, Hong Kong SAR, China (e-mail: whyeung@ie.cuhk.edu.hk). %This work was funded in part by the NSFC (Grant No. 61771259) and the Fundamental Research Funds for the Central Universities, Nankai University.}}

%Email: pertox4726@gmail.com
%Email: 09blueleaves09@gmail.com

\maketitle

\begin{abstract}
In this paper, we put forward secure network function computation over a directed acyclic network. In such a network, a sink node is required to compute with zero error a target function of which the inputs are generated as source messages at multiple source nodes, while a wiretapper, who can access any one but not more than one wiretap set in a given collection of wiretap sets, is not allowed to obtain any information about a security function of the source messages. The secure computing capacity for the above model is defined as the maximum average number of times that the target function can be securely computed with zero error at the sink node with the given collection of wiretap sets and security function for one use of the network. The characterization of this capacity is in general overwhelmingly difficult. In the current paper, we consider securely computing linear functions with a wiretapper who can eavesdrop any subset of edges up to a certain size~$r$, referred to as the {\em security level}, with the security function being the identity function. We first prove an upper bound on the secure computing capacity, which is applicable to arbitrary network topologies and arbitrary security levels. This upper bound depends on the network topology and security level. Furthermore, we obtain an upper bound and a lower bound on this bound, which are both in closed form. In particular, when the security level~$r$ is equal to $0$, our upper bound reduces to the computing capacity without security consideration. Also, we discover the surprising fact that for some models, there is no penalty on the secure computing capacity compared with the computing capacity without security consideration. Furthermore, we obtain an equivalent expression of the upper bound by using a graph-theoretic approach, and accordingly we develop an efficient approach for computing this bound.
%By applying our upper bound, we obtain a non-trivial upper bound on the maximum security level such that the function can be securely computed with a positive rate.
On the other hand, we present a construction of linear function-computing secure network codes and obtain a lower bound on the secure computing capacity. By our code construction, for the linear function which is over a given finite field, we can always construct a (vector-) linear function-computing secure network code over the same field. We also give some sufficient conditions for the tightness of the lower bound in terms of the network topology. With this lower bound and the upper bound we have obtained, the secure computing capacity for some classes of secure models can be fully characterized.
Another interesting case is that the security function is the same as the target function, which will be investigated in Part~\Rmnum{2} of this paper.
\end{abstract}

% Note that keywords are not normally used for peer review papers.
%\begin{IEEEkeywords}
%Network coding, network error correction, linear multicast/broadcast/dispersion, generic network codes, linear network error correction multicast/broadcast/dispersion/generic codes, Singleton bound,  multicast/broadcast/dispersion/generic MDS codes, constructive algorithms.
%\end{IEEEkeywords}

\IEEEpeerreviewmaketitle

%%%%%%%%%%%%%%%%%%%%%%%%%%%%%%%%%-------Introduction------%%%%%%%%%%%%%%%%%

\section{Introduction}

In this paper, we investigate \textit{secure network function computation} that incorporates information-theoretic security with zero-error network function computation. A general setup of the model with the single destination is presented as follows. In a directed acyclic graph $\mG$, \rmnum{1}) the single sink node~$\rho$ is required to compute repeatedly (with zero error) a function $f$, called the {\em target function}, whose arguments are source messages generated at a set of source nodes $S$; and \rmnum{2}) a wiretapper, who can access any one but not more than one edge subset $W\in \mW$, is not allowed to obtain any information about another function~$\zeta$, called the {\em security function}, whose arguments are also the source messages. Here, $W$ and $\mW$ are referred to as the \textit{wiretap set} and the collection of wiretap sets, respectively. The graph $\mG$, together with $S$ and $\rho$, forms a \emph{network} $\mN$, and we use the quadruple $(\mN,f,\mW,\zeta)$ to denote this model of secure network function computation.

We note that when both the target function $f$ and the security function $\zeta$ are the identity function, the secure model $(\mN,f,\mW,\zeta)$ degenerates to a secure network coding model (cf.~\cite{secure-conference,Cai-Yeung-SNC-IT,Rouayheb-IT,GY-SNC-Reduction,Silva-UniversalSNC,GYF-LEP-SNC}). When no security constraint is considered, i.e., $\mW=\emptyset$, the secure model $(\mN,f,\mW,\zeta)$ degenerates to the model of network function computation (cf.~\cite{Koetter-CISS2004,Guang_NFC_TIT19,Appuswamy11,Appuswamy13,Appuswamy14,HTYGuang-TIT18,
Ramamoorthy-Langberg-JSAC13-sum-networks, Rai-Dey-TIT-2012}).

\subsection{Related Works}

{\em Information-theoretic security} dates from Shannon's celebrated paper~\cite{Shannon-secrecy}, in which the well-known \textit{Shannon cipher system} was studied. In this system, a sender wishes to transmit a private message to a receiver via a ``public'' channel which is eavesdropped by a wiretapper, and it is required that this wiretapper cannot obtain any information about the private message. For this purpose, the sender applies a random key to encrypt the message and then transmit this encrypted message via the ``public'' channel. The random key is shared with the receiver via a ``secure'' channel that is inaccessible by the wiretapper. The receiver can recover the private message from the encrypted message and the random key, while the wiretapper cannot obtain any information about the private message. Another well-known cipher system of information-theoretic security is \textit{secret sharing}, proposed independently by Blakley \cite{Blakley_secret-sharing-1979} and Shamir \cite{Shamir_secret-sharing-1979}. In this system, a secret is encoded into shares which are distributed among a set of participants, and only the qualified subsets of participants can recover the secret, while no information at all about the secret can be obtained from the shares of any unqualified set of participants. The Shannon cipher system can be regarded as a special case of secret sharing. \textit{Wiretap channel~\Rmnum{2}}, proposed by Ozarow and Wyner \cite{wiretap-channel-II}, is a related system of information-theoretic security, in which the sender needs to transmit the private message to the receiver through a set of noiseless point-to-point channels without leaking any information about the private message to a wiretapper who can fully access any one but not more than one subset of the channels up to a certain size. Wiretap channel~\Rmnum{2} can be regarded as a special case of secret sharing.

In the paradigm of network coding, the information-theoretically secure problem in the presence of a wiretapper, called {\em secure network coding}, was introduced by Cai and Yeung in \cite{secure-conference, Cai-Yeung-SNC-IT}. In the wiretap network model of secure network coding, the source node multicasts the source message to all the sink nodes which as legal users are required to decode the source message with zero error; while the wiretapper, who can access any one wiretap set of edges, is not allowed to obtain any information about the source message. Logically, the foregoing three classical information-theoretically secure systems, the Shannon cipher system, secret sharing and wiretap channel~\Rmnum{2} can be formulated as special cases of the wiretap network model of secure network coding.

{\em Network function computation} was considered in the literature \cite{Appuswamy11,Appuswamy13,Appuswamy14,HTYGuang-TIT18,Guang_NFC_TIT19}. In the model with a single destination, the sink node is required to repeatedly compute with zero error a target function of the source messages generated at multiple source nodes over a directed acyclic network. When the sink node is required to compute the identity function of the source messages, or equivalently, recover the source messages, the problem degenerates to the network coding problem \cite{Zhang-book, Yeung-book, Fragouli-book, Fragouli-book-app, Ho-book}. Appuswamy~\textit{et~al.}~\cite{Appuswamy11} investigated the fundamental {\em computing capacity}, i.e., the maximum average number of times that the function can be computed with zero error for one use of the network, and gave a cut-set based upper bound that is valid under certain constraints on either the network topology or the target function. Huang~\textit{et~al.}~\cite{HTYGuang-TIT18} obtained an enhancement of Appuswamy~\textit{et~al.}'s upper bound that can be applied for arbitrary functions and arbitrary network topologies.
For the special cases, e.g., computing an arbitrary function over a multi-edge tree network and computing the identity function or the algebraic sum function over an arbitrary network topology, the upper bounds in \cite{Appuswamy11} and \cite{HTYGuang-TIT18} coincide and are tight. Guang~\textit{et~al.}~\cite{Guang_NFC_TIT19} proved an improved upper bound on the computing capacity by using a novel approach based on cut-set strong partition. This improved upper bound is applicable to arbitrary target functions and arbitrary network topologies. In particular, the improved upper bound is asymptotically achievable for all previously considered network function computation problems whose computing capacities are known.

\subsection{Contributions and Organization of the Paper}

For the model of secure network function computation $(\mN,f,\mW,\zeta)$, we consider in this paper the special case that the security function $\zeta$ is the identity function, namely that we need to protect the information sources from being leaked to the wiretapper. We use $(\mN, f, \mW)$ to denote this special case of the model. The notion of security is considered in almost all previously studied information-theoretic security models, e.g., the Shannon cipher system~\cite{Shannon-secrecy}, the secret sharing~\cite{Blakley_secret-sharing-1979,Shamir_secret-sharing-1979}, the wiretap channel~\Rmnum{2}~\cite{wiretap-channel-II}, and secure network coding~\cite{Cai-Yeung-SNC-IT,Rouayheb-IT}. Another interesting case is that the security function is the same as the target function, i.e., $\zeta=f$. This will be treated in Part~\Rmnum{2} of the current paper.

From the information theoretic point of view, we are interested in characterizing the secure computing capacity for $(\mN,f,\mW)$, which is defined as the maximum average number of times that the function $f$ can be securely computed with zero error at the sink node with the given collection of wiretap sets $\mW$ for one use of the network $\mN$. However, characterizing this secure computing capacity for an arbitrary security model $(\mN,f,\mW)$ is overwhelmingly difficult and complicated, and even for the simpler case of network function computation without any security consideration, the capacity characterization is still open~\cite{Guang_NFC_TIT19}. Thus, in the paper we focus on the model of securely computing a linear function over a finite field on an arbitrary network where the wiretapper can eavesdrop any one edge subset up to a size $r$, referred to as the {\em security level}. We note that the linear functions are not only an important class of target functions but also the only class of non-identity functions whose computing capacities can be characterized without any security consideration.

In this paper, we first prove a non-trivial upper bound on the secure computing capacity, which is applicable to arbitrary network topologies and arbitrary security levels. This upper bound improves the upper bound we previously obtained in~\cite{GBY-SecureNFC-ISIT21}. In particular, when no security is considered, i.e., $r=0$, our upper bound reduces to the capacity for computing a linear function over the network without security consideration, which we call the \textit{computing capacity} for simplicity. Furthermore, we discover that for some models, there is no penalty on the secure computing capacity compared with the computing capacity without security consideration. In other words, the secure computing capacity may coincide with the computing capacity for a security level $r>0$. From the upper bound on the secure computing capacity we have obtained, we also obtain a non-trivial upper bound on the maximum security level such that the function can be securely computed with a positive rate.

Our upper bound on the secure computing capacity, which depends on the network topology and security level, is graph-theoretic and not in closed form. In order to compute this upper bound efficiently, we prove an equivalent expression by using a graph-theoretic approach. Accordingly, we devise an algorithm for computing this bound whose computational complexity is in a linear time of the number of edges in the network. We further prove an upper bound and a lower bound on this upper bound on the secure computing capacity, which are both in closed form.

Furthermore, we put forward a construction of linear (function-computing) secure network codes and thus obtain a lower bound on the secure computing capacity. By our code construction, with the target function being a linear function over a given finite field, we can always construct a (vector-) linear secure network code over the same field with rate up to the obtained lower bound on the secure computing capacity. Besides, we give two sufficient conditions for the tightness of the lower bound in terms of the network topology. With this lower bound and the upper bound we have obtained, we can fully characterize the secure computing capacity for some classes of secure models. Finally, an example is given that not only illustrates our code construction but also shows the existence of a linear secure network code that cannot be obtained by our code construction.

The paper is organized as follows. In Section~\ref{model_SNFC}, we formally present the model of secure network function computation and function-computing secure network coding. The upper bound is proved in Section~\ref{sec:UppBound}, followed by an efficient graph-theoretic approach for computing the upper bound. Section~\ref{sec:LinearSNC} is devoted to linear secure network coding for the model of secure network function computation, including the lower bound on the secure computing capacity, the code construction, the verification of the computability and security conditions, and the upper bounds on the minimum required field size for the code construction. In Section~\ref{sec:concl}, we conclude with a summary of our results and a remark on future research.

\section{Preliminaries}\label{model_SNFC}

\subsection{Model of Secure Network Function Computation}

Let $\mG=(\mV,\mE)$ be a directed acyclic graph, where $\mV$ and $\mE$ are a finite set of nodes and a finite set of edges, respectively. We allow multiple edges between two nodes and assume that a symbol taken from a finite alphabet $\mB$ can be reliably transmitted on each edge for each use, i.e., we take the capacity of each edge to be $1$ with respect to the alphabet $\mB$. For an edge $e \in \mE$, the \emph{tail} node and \emph{head} node of $e$ are denoted by $\tail(e)$ and $\head(e)$, respectively. For a node $u \in \mV$, we let $\ein(u)=\{ e \in \mE:\head(e)=u \}$ and $\eout(u)=\{ e \in \mE:\tail(e)=u \}$.

In the graph $\mG$, if a sequence of edges $(e_1,e_2,\cdots,e_m)$ satisfies $\tail(e_1)=u$, $\head(e_m)=v$ and $\tail(e_i)=\head(e_{i-1})$ for $i=2,3,\cdots,m$, then $(e_1,e_2,\cdots,e_m)$ is called a {\em path} from the node $u$ (or the edge $e_1$) to the node $v$ (or the edge $e_m$). In particular, a single edge $e$ is regarded as a path from $\tail(e)$ to $\head(e)$ (or from $e$ to itself). Furthermore, we write $d\rightarrow e$ (or $\tail(d) \rightarrow \head(e)$, $\tail(d)\rightarrow e$, $d\rightarrow \head(e)$) if there exists a path from edge $d$ to edge $e$.
We now consider two disjoint subsets of nodes $U$ and $V$. An edge subset $C\subseteq \mE$ is called a \emph{cut} separating $V$ from $U$ if for each pair $(u,v)$ of $u \in U$ and $v \in V$, no path exists from $u$ to $v$ upon removing the edges in $C$. In particular, if $U=\{ u \}$ and $V=\{ v \}$ are two disjoint singleton subsets of nodes, a cut separating $V$ from $U$ is called a cut separating node $v$ from node $u$. The \emph{capacity} of a cut separating $V$ from $U$ is defined as the size of this cut. A cut $C$ separating $V$ from $U$ is called a \emph{minimum cut} separating $V$ from $U$ if there does not exist a cut $C'$ separating $V$ from $U$ such that $\vert C' \vert < \vert C \vert$. The capacity of a minimum cut separating $V$ from $U$ is called the \emph{minimum cut capacity} separating $V$ from $U$, denoted by $\mincut(U,V)$. In particular, when $U=\{ u \}$ and $V=\{ v \}$ with $u \neq v$, the minimum cut capacity separating $v$ from $u$ is denoted by $\mincut(u,v)$.

%{\color{red}(DELETE?) If an edge set $A\subseteq E$ is a cut separating a node $v$ (resp. a set of nodes $\widehat{V}$ and a set of edges $\xi$) from another node $u$, then we say that the edge set $A$ {\em separates} $v$ (resp.~$\widehat{V}$ and $\xi$) {\em from} $u$. Note that if $A$ separates $v$ (resp. $\widehat{V}$ and $\xi$) from $u$, then every path from $u$ to $v$ (resp. $\widehat{V}$ and $\xi$) passes through at least one edge in $A$.}

We let $S \subset \mV$ be the set of \emph{source nodes} $\sigma_1, \sigma_2, \cdots, \sigma_s$ and $\rho \in \mV \setminus S$ be the single \emph{sink node}, where each source node $\sigma_i$ has no input edges and the single sink node $\rho$ has no output edges, i.e., $\ein(\sigma_i)=\eout(\rho)=\emptyset$, $i=1,2,\cdots,s$. We further assume that there exists a directed path from every node $u\in \mathcal{V}\setminus \{\rho\}$ to $\rho$ in $\mG$. Then, the source nodes are all the nodes in $\mV$ without input edges and the sink node $\rho$ is the unique  node in $\mV$ without output edges. The graph $\mG$, together with $S$ and $\rho$, forms a \emph{network} $\mN$, i.e., $\mN=(\mG, S, \rho)$.

Let $f:~\prod_{i=1}^s\mA_i\to \mO$ be a nonconstant function, called the \emph{target function}, that is needed to be computed with zero error at the sink node $\rho$, where $\mA_i$, $1\leq i \leq s$ and $\mO$ are all finite alphabets. %\footnote{We only consider nonconstant functions as target functions in the paper, because any transmission and computation is unnecessary for computing a constant function over a network.}
Further, we assume without loss of generality that the $i$th argument of the target function $f$ is generated at the $i$th source node $\sigma_i$ for $i=1,2,\cdots,s$. Let $\ell$ and $n$ be two positive integers. We consider computing the target function $\ell$ times by using the network $n$ times, i.e., by transmitting at most $n$ symbols in $\mB$ on each edge in $\mE$. For each $i=1,2,\cdots,s$, we let the {\em information source} at the $i$th source node $\sigma_i$ be a random variable $M_i$ according to the uniform distribution on $\mA_i$. All the sources $M_i$, $1\leq i \leq s$ are mutually independent. Let $M_S=(M_{1},M_{2}, \cdots, M_s)$. The $i$th source node $\sigma_i$ sequentially generates $\ell$ independent identical distributed (i.i.d.) random variables $M_{i,1}, M_{i,2}, \cdots, M_{i,\ell}$ with generic random variable $M_i$. We let $\vM_i=(M_{i,1},M_{i,2}, \cdots, M_{i,\ell})$, called the \emph{source message generated by~$\sigma_i$}. We further let $\vM_S=(\vM_1,\vM_2,\cdots,\vM_s)$ be the \emph{source message vector generated by $S$}. The $\ell$ values of the target function $f$
\begin{align}\label{function_f}
f(\vM_S)\triangleq\big(f(M_{1,j},M_{2,j}, \cdots, M_{s,j}): j=1,2,\cdots,\ell \big)
\end{align}
are required to be computed at $\rho$ with zero error.

In addition, we consider a collection of edge subsets $\mW$ where each edge subset $W \in \mW$ is called a \emph{wiretap set}, and another nonconstant function $\zeta: \prod_{i=1}^{s}\mA_i \rightarrow \mathcal{Q}$, called the {\em security function}, where $\mathcal{Q}$ is a finite alphabet. In this model, when the $\ell$ target function values $f(\vM_S)$ are computed at $\rho$ through the network $\mN$, the $\ell$ values of the security function $\zeta$
\begin{align}\label{def_zeta_MS}
  \zeta(\vM_S) \triangleq \big( \zeta(M_{1, j}, M_{2, j}, \cdots, M_{s, j}): ~j=1, 2, \cdots, \ell \big)
\end{align}
are required to be protected from a wiretapper who can access any one but not more than one wiretap set $W \in \mW$. The collection of wiretap sets $\mW$ and the security function $\zeta$ are known by the source nodes and the sink node but which wiretap set in $\mW$ is eavesdropped by the wiretapper is unknown. We have completed the specification of our model of secure network function computation, denoted by $(\mN,f,\mW,\zeta)$.

In this paper, we consider a special case of our model that the security function $\zeta$ is the identity function, namely that we need to protect the information sources from being leaked to the wiretapper. We use $(\mN, f, \mW)$ to denote this special case of the model. The notion of security is considered in almost all previously studied information-theoretic security models, e.g., the Shannon cipher system~\cite{Shannon-secrecy}, the secret sharing~\cite{Blakley_secret-sharing-1979,Shamir_secret-sharing-1979}, the wiretap channel~\Rmnum{2}~\cite{wiretap-channel-II}, and secure network coding~\cite{Cai-Yeung-SNC-IT}. In the next subsection, we will define a function-computing secure network code on $(\mN,f,\mW)$.

Another interesting case of the model of secure network function computation is that the security function is the same as the target function, i.e., $\zeta=f$. This case will be treated in the next paper.

\subsection{Function-Computing Secure Network Code}

In order to combat the wiretapper in the secure model $(\mN, f, \mW)$, it is necessary to randomize the source messages, which is similar to the previously information-theoretic security models (e.g., the Shannon cipher system~\cite{Shannon-secrecy}, the secret sharing~\cite{Blakley_secret-sharing-1979,Shamir_secret-sharing-1979}, the wiretap channel~\Rmnum{2}~\cite{wiretap-channel-II}, and secure network coding~\cite{Cai-Yeung-SNC-IT}). The reason is as follows. Suppose no randomness is used to randomize the source messages. Then, the message transmitted on each edge is a function of the source message vector $\vM_S$ and hence is not independent of $\vM_S$ unless the transmitted message takes a constant value, which is equivalent to transmitting nothing on that edge. If this is the case, nothing can be computed at the sink node. Therefore, randomness is necessary for transmitting a message securely.

As part of the network code to be defined, we assume that for $i=1,2,\cdots, s$, a random variable $\vK_i$, called a \textit{key}, which is distributed uniformly on a finite set $\mK_i$, is available to the source node $\sigma_i$. We let $\vK_S=(\vK_1,\vK_2,\cdots,\vK_s)$. Further, assume that all the keys $\vK_i$ and the source messages $\vM_i$, $i=1,2,\cdots, s$ are mutually independent.

Now, we consider securely computing the target function $f$ $\ell$ times by using the network $n$ times under the collection of wiretap sets $\mW$. An $(\ell,n)$ {\em (function-computing) secure network code} for $(\mN, f, \mW)$ is defined as follows. First, we let $\vm_i \in \mA_i^{\ell}$ and $\vk_i\in \mK_i$ be arbitrary outputs of the source message $\vM_i$ and the key $\vK_i$, respectively, for $i=1,2,\cdots,s$. Accordingly, let $\vm_S = (\vm_1,\vm_2,\cdots,\vm_s)$ and $\vk_S = (\vk_1,\vk_2,\cdots,\vk_s)$, which can be regarded as two arbitrary outputs of $\vM_S$ and $\vK_S$, respectively. An $(\ell,n)$ secure network code $\hmbC$ consists of a {\em local encoding function} $\htheta_e$ for each edge $e \in \mE$, where
\begin{equation}\label{snc_local}
\htheta_e:
\begin{cases}
\qquad \mA_i^\ell \times \mK_i \rightarrow \mB^n , & \text{if } \tail(e) =\sigma_i \text{ for some } i,\\
\prod\limits_{d \in \ein(\tail(e))} \mB^n \rightarrow \mB^n , & \text{otherwise;}
\end{cases}
\end{equation}
and a decoding function
$
\hvarphi:~\prod_{\ein(\rho)}\mB^n \rightarrow \mathcal{O}^{\ell}
$
at the sink node $\rho$, which is used to compute the target function $f$ with zero error. Furthermore, let $\vy_e\in \mB^n$ be the message transmitted on each edge $e\in \mE$ by using the code $\hmbC$ under the source message vector $\vm_S$ and the key vector $\vk_S$. With the encoding mechanism as described in~\eqref{snc_local}, we readily see that $\vy_e$ is a function of $\vm_S$ and $\vk_S$, denoted by $\hg_e(\vm_S, \vk_S)$ (i.e., $\vy_e=\hg_e(\vm_S, \vk_S)$), where $\hg_e$ can be obtained by recursively applying the local encoding functions $\htheta_e$, $e\in \mE$. More precisely, for each $e\in \mE$,
\begin{equation*}
\hg_{e}(\vm_S,\vk_S) =
\begin{cases}
\htheta_{e}(\vm_i,\vk_i), & \text{if } \tail(e)=\sigma_i \text{ for some } i,\\
\htheta_{e}\big(\hg_{\ein(u)}(\vm_S,\vk_S)\big), & \text{otherwise},
\end{cases}
\end{equation*}
where $u=\tail(e)$ and $\hg_E(\vm_S,\vk_S)=\big(\hg_{e}(\vm_S,\vk_S):~e \in E\big)$ for an edge subset $E \subseteq \mE$ (in particular, $\hg_{\ein(u)}(\vm_S,\vk_S)=\big(\hg_{e}(\vm_S,\vk_S):~e \in \ein(u)\big)$).
We call $\hg_{e}$ the \emph{global encoding function} of the edge $e$ for the code $\hmbC$.

For the secure model $(\mN,f,\mW)$, we say an $(\ell,n)$ secure network code $\hmbC=\big\{ \htheta_{e}:~e\in \mE;~  \hvarphi \big\}$ is {\em admissible} if the following {\em computability} and {\em security conditions} are satisfied:
\begin{itemize}
  \item \textbf{\em computability condition}: the sink node $\rho$ computes the target function $f$ with zero error, i.e., for all $\vm_S\in \prod_{i=1}^s\mA_i^{\ell}$ and $\vk_S\in \prod_{i=1}^{s}\mK_i$,
  \begin{align}\label{dc}
  \hvarphi \big(\hg_{\ein(\rho)}(\vm_S,\vk_S) \big)=f(\vm_S);
  %, \qquad \forall~\vm_S\in \mA^{s\cdot\ell} ~\text{ and }~ \vk_S\in \prod_{i=1}^{s}\mK_i;
  \end{align}
  \item \textbf{\em security condition}: for any wiretap set $W\in \mW$, $\vY_W$ and $\vM_S$ are independent, i.e.,
      \begin{align}\label{sc}
      I(\vY_W ; \vM_S)=0,
      \end{align}
      where $\vY_W=(\vY_e: e\in W)$, and $\vY_e \triangleq \hg_{e}(\vM_S,\vK_S)$ is the random variable transmitted on the edge~$e$.
  \end{itemize}
The \emph{secure computing rate} of such an admissible $(\ell, n)$ secure network code $\hmbC$ is defined by
%\begin{align*}
%R(\hmbC) \triangleq \frac{H(f(\vM_S))}{n\cdot\log|\mathcal{B}|}
%=\frac{\ell \cdot H(f(M_S))}{n\cdot\log|\mathcal{B}|}.
%\end{align*}
%We note that $\frac{H(f(M_S))}{\log|\mathcal{B}|}$ is constant for the given target $f$ and alphabet $\mB$. So it is equivalent to defining the secure computing rate by
\begin{align*}
R(\hmbC) \triangleq\frac{\ell}{n},
\end{align*}
i.e., the average number of times the function $f$ can be securely computed with zero error at $\rho$ under the collection of wiretap sets $\mW$ for one use of the network $\mN$. %If there exists an admissible $(\ell, n)$ secure network code, the secure computing rate $\ell/n$ is called {\em achievable}.
Further, we say that a nonnegative real number $R$ is {\em achievable} if $\forall~\epsilon > 0$, there exists an admissible $(\ell, n)$ secure network code $\hmbC$ such that
$$R(\hmbC)=\frac{\ell}{n}>R-\epsilon.$$
We readily see that the secure computing rate $\ell/n$ of an admissible $(\ell, n)$ secure network code must be achievable. The {\em secure computing rate region} for the secure model $(\mN,f,\mW)$ is defined as
\begin{align}\label{def:secure_rate_region}
\mathfrak{R}(\mN,f,\mW) \triangleq \Big\{ R:~R \text{ is achievable for $(\mN,f,\mW)$} \Big\},
\end{align}
which is evidently closed and bounded. Consequently, the {\em secure computing capacity} for $(\mN, f, \mW)$ is defined as
\begin{align}\label{defi_secure_CR}
\hmC(\mN,f,\mW)\triangleq  \max~\mathfrak{R}(\mN, f, \mW).
\end{align}

From the information theoretic point of view, we are interested in the secure computing capacity $\hmC(\mN,f,\mW)$ for an arbitrary model $(\mN,f,\mW)$. However, characterizing this capacity with the general setup is overwhelmingly difficult and complicated, and even for the simpler case of network function computation without any security consideration, the computing capacity characterization is still open (cf.~\cite{Guang_NFC_TIT19}). In fact, for computing an arbitrary target function over an arbitrary network without any security consideration, characterizing the computing capacity is in general difficult. So far, only the computing capacities for linear functions over finite fields and for identity functions have been fully characterized. Note that computing the identity function degenerates to the regular network coding problem. As secure network function computation is a generalization of network function computation, we naturally consider secure network computing for the target functions whose computing capacities have been determined when no security is considered. Thus, in this paper we focus on linear functions over finite fields. This is not only an important class of target functions but also the only non-trivial class of functions whose computing capacities without any security consideration can be determined. To be specific, the target function $f$ is a \emph{linear function} over a finite field $\Fq$ of form
\begin{align*}
f(m_1, m_2, \cdots, m_s)=\sum_{i=1}^s a_i \cdot m_i,
\end{align*}
where $m_i,\,a_i\in \Fq$ for all $i=1,2,\cdots,s$~(i.e., $\mA_i=\mO=\Fq$, $\forall~1\leq i \leq s$ and $f:~\Fq^s \rightarrow \Fq$). Accordingly, all the independent information sources $M_i$, $i=1,2,\cdots,s$ are random variables distributed uniformly on the finite field $\Fq$. Moreover, we consider a special collection of wiretap sets
\begin{align*}
\mW_r\triangleq \big\{ W \subseteq \mE:~0 \leq \vert W \vert \leq r \big\},
\end{align*}
i.e., the wiretapper can eavesdrop any one subset of edges in the network up to a size $r$, which is referred to as the {\em security level}. We remark that the empty set $\emptyset$, regarded as the wiretap set of size $0$, is in $\mW_r$. For notational simplicity, we write the secure model $(\mN,f,\mW_r)$ and secure computing capacity $\hmC(\mN,f,\mW_r)$ as $(\mN,f,r)$ and $\hmC(\mN,f,r)$, respectively. In the rest of the paper, we assume without loss of generality that $\mB=\Fq$, i.e., an element in the field $\Fq$ can be reliably transmitted on each edge for each use.

%\begin{example}\label{exam_secure}
%\end{example}

\section{Upper Bound on the Secure Computing Capacity}\label{sec:UppBound}

We first observe that the characterization of secure computing capacity for a linear function over a finite field is equivalent to the characterization of secure computing capacity for an algebraic sum over the same field. This is explained as follows. We consider the model of secure network function computation $(\mN,f,r)$ with $f$ being a linear function over a finite field $\Fq$, i.e.,
\begin{align*}
f(m_1,m_2,\cdots,m_s)=\sum_{i=1}^{s}a_i \cdot m_i.
\end{align*}
For the indices $i$ with $a_i=0$, we remove the terms $a_i \cdot m_i$ from $f$ to form a new linear function $f'$, and at the same time remove the source nodes $\sigma_i$ together with the output edges in $\eout(\sigma_i)$ from the graph~$\mG$ to form a new graph $\mG'$. We update the network $\mN$ to a new one $\mN'=\big( \mG', S\setminus \{ \sigma_i:~a_i=0 \}, \rho \big)$. We readily see that with the security level $r$, securely computing $f$ over $\mN$ is equivalent to securely computing $f'$ over $\mN'$, and $\hmC(\mN,f,r)=\hmC(\mN',f',r)$. Thus, we can assume without loss of generality that in the model $(\mN,f,r)$ with $f(m_1,m_2,\cdots,m_s)=\sum_{i=1}^{s}a_i \cdot m_i$, we have $a_i \neq 0$ for all $1\leq i \leq s$. Now, let $x_i=a_i \cdot m_i$ and consider the algebraic sum $g(x_1,x_2,\cdots,x_s)=\sum_{i=1}^{s}x_i$ over $\Fq$.
Since $a_i \neq 0$ for all $1\leq i \leq s$, $m_i$ and $x_i$ can be determined from each other, and hence an admissible $(\ell,n)$ secure network code for $(\mN,f,r)$ can be readily modified into an admissible $(\ell, n)$ secure network code for $(\mN,g,r)$, and vice versa. So, we have $\hmC(\mN,f,r)=\hmC(\mN,g,r)$. By the above discussion, we can see that in order to investigate the secure network function computation for a linear function, it suffices to consider secure network function computation for an algebraic sum. Accordingly, in the rest of the paper, we let the target function $f$ be the algebraic sum over a finite field $\Fq$, i.e.,
\begin{align}\label{equ-alge_sum}
f(m_1,m_2,\cdots,m_s)=\sum_{i=1}^{s} m_i.
\end{align}

\subsection{The Upper Bound}

Next, we present some graph-theoretic notations. %{\color{red}For two nodes $u$ and $v$ in the graph $\mG$, we write $u\rightarrow v$ if there exists a path from $u$ to $v$. If no path exists from $u$ to $v$, we say that $v$ is \emph{separated} from $u$.}
Given a set of edges $C\subseteq \mathcal{E}$, we define three subsets of the source nodes as follows:
\begin{align*}
 D_C & =  \big\{ \sigma\in S:~\exists\ e\in C \text{ s.t. } \sigma\rightarrow e \big\},\\
 I_C & = \big\{ \sigma\in S:~ \sigma \nrightarrow \rho \text{ upon deleting the edges in $C$ from $\mathcal{E}$} \big\},\\
 J_C & = D_C \setminus  I_C,
\end{align*}
where $\sigma \nrightarrow \rho$ denotes that there exists no path from $\sigma$ to $\rho$. Since we assume that
there is a directed path from every node $u\in \mathcal{V}\setminus \{\rho\}$ to~$\rho$, in particular, a directed path from every source node $\sigma \in S$ to~$\rho$, we see that $I_C\subseteq D_C$. Here, $J_C$ is the subset of the source nodes $\sigma$ satisfying that there exists not only a path from $\sigma$ to $\rho$ passing through an edge in $C$ but also a path from $\sigma$ to $\rho$ not passing through any edge in~$C$.\footnote{Similarly, since it is assumed that there is a directed path from every node $u\in \mathcal{V}\setminus \{\rho\}$ to $\rho$, there is a path from $\sigma$ to $\rho$ passing through an edge $e\in C$ provided that there is a path from $\sigma$ to $e$.} In the network~$\mathcal{N}$, an edge set $C$ is said to be a {\em cut set} if $I_C\neq \emptyset$, and we let $\Lambda(\mathcal{N})$ be the family of all the cut sets, i.e.,
\begin{align*}
\Lambda(\mathcal{N})=\big\{ C\subseteq \mE:\ I_C \neq \emptyset \big\}.
\end{align*}
In particular, we say a cut set $C$ is a {\em global cut set} if $I_C=S$. Now, we present an upper bound on the secure computing capacity $\hmC(\mN,f,r)$ in the following theorem.

\begin{thm}\label{thm_upper_capa}
Consider a model of secure network function computation $(\mN, f, r)$, where the target function~$f$ is an algebraic sum over a finite field $\Fq$. Then,
\begin{align}\label{equ_upper_bound}
   \hmC(\mN,f,r) \leq
   \min_{ (W,C)\in \mW_r \times \Lambda(\mN):\atop  W\subseteq C \text{ \rm  and }  D_W \subseteq  I_C} \big( |C|-|W| \big).
  %  \min_{\color{blue}\text{ \rm all pairs } (W,C):\atop W\in \hmW_r, C \in\Lambda(\mN) \text{ \rm and } W\subseteq C} |C|-|W|.
    %\\
   %\min_{W \in \hmW_r} \min_{C \in\Lambda(\mN) \atop \text{s.t. } C\supseteq W} |C|-|W|\\
   % \min_{0\leq R \leq r} \min_{\text{ \rm all pairs } (W,C):\atop W\in \hmW_R, C \in\Lambda(\mN) \text{ \rm and } W\subseteq C} |C|-R,
   \end{align}
%where
%\begin{align*}
%\hmW_r \triangleq \big\{ W \subseteq \mE:~ 0\leq \vert W \vert \leq r \text{ and } J_W=\emptyset \big\}\subseteq \mW_r.\footnotemark
%\end{align*}
\end{thm}

%Here, we remark that for $R=0$, $\hmW_0$ is the singleton that contains the wiretap set of size $0$, i.e., the empty set $\emptyset$.}

\begin{IEEEproof}
We let $\hmbC$ be an arbitrary admissible $(\ell,n)$ secure network code for the secure model $(\mN, f, r)$, of which all the global encoding functions are $\hg_e$, $e\in\mE$. We consider an arbitrary cut set $C \in \Lambda(\mN)$, i.e., $I_C\neq \emptyset$. Recall that all the source messages $M_{i,j}$ for $i=1,2,\cdots, s$ and $j=1,2,\cdots,\ell$ are i.i.d. according to the uniform distribution on $\Fq$. Thus, we have
\begin{align}\label{equ1}
H\Big(\sum_{i\in I_C}\vM_i\Big)=\ell \cdot H\Big( \sum_{i\in I_C} M_{i} \Big)=\ell \cdot \log q,\footnotemark
\end{align}
where we recall that $\vM_i$ is the vector of the $\ell$ random variables $M_{i,j}$, $1\leq j \leq \ell$ with generic random variable $M_i$, i.e., $\vM_i=(M_{i,1},M_{i,2}, \cdots, M_{i,\ell})$, and similar to \eqref{function_f}, we let
\begin{align*}
\sum_{i\in I_C}\vM_i\triangleq \Big(\sum_{i\in I_C}M_{i,j} : j=1,2,\cdots,\ell \Big).
\end{align*}
\footnotetext{We remark that for a subset of the source nodes $I\subseteq S$, we also use $I$ to represent the index set $\{i:~\sigma_i\in I\}$. This abuse of notation should cause no ambiguity.}

%By the independence of $\vM_1,\vM_2,\cdots,\vM_s$, we further have
%\begin{align}\label{equ2}
%\ell\cdot\log q = H(\sum_{i\in I_C}\vM_i)=H(\sum_{i\in I_C}\vM_i|\vM_{S\setminus I_C}),
%\end{align}
%where the index set $S\setminus I_C \triangleq \{i:~1\leq i \leq s \text{ and } i\in I_C\}$.

Now, we consider an arbitrary wiretap set $W \subseteq C$ that satisfies $0\leq |W| \leq r$ and $D_W\subseteq I_C$. We remark that the empty set is also such a wiretap set. By the security condition~\eqref{sc}, we have
\begin{align*}
  H(\vM_S)=H(\vM_S|\vY_W),
\end{align*}
which implies that
\begin{align}\label{eq_Omin_2}
  H(\vM_{I_C})=H(\vM_{I_C}|\vY_W),
\end{align}
where we let $\vM_{I_C} \triangleq \big(\vM_i:~i\in I_C\big)$.
It follows from \eqref{eq_Omin_2} that
%\begin{align}
%  H\left(\sum_{i\in I_C}\vM_i\right)=H\left(\sum_{i\in I_C}\vM_i\bigg|\vY_W\right),
%\end{align}
\begin{align}
H\bigg(\sum_{i\in I_C}\vM_i\bigg)=H\bigg(\sum_{i\in I_C}\vM_i \Big|\vY_W\bigg),
\end{align}
or equivalently,
\begin{align}\label{pf-equ1}
  H\bigg(\sum_{i\in I_C}\vM_i \bigg)+H(\vY_W)=H\bigg(\sum_{i\in I_C}\vM_i, \vY_W\bigg),
\end{align}
because $\sum_{i\in I_C}\vM_i$ is a function of $\vM_{I_C}$.

By the definition of $D_W$, there exists no path from any source node $\sigma\in S \setminus D_W$ to any edge in $W$. Then, we see that \rmnum{1}) $\vY_W$ is independent of $\vM_{S \setminus D_W}$ and $\vK_{S \setminus D_W}$; \rmnum{2)} $\vY_W=\hg_{W}\big(\vM_S,\vK_S \big)$ depends only on $\big(\vM_{D_W},\vK_{D_W} \big)$. Following \rmnum{2}), we write
\begin{align}\label{pf-equ11}
\vY_W=\hg_{W}\big(\vM_S,\vK_S \big)=\hg\,'_{W}\big(\vM_{D_W},\vK_{D_W} \big).
\end{align}
Since $D_W\subseteq I_C$, $\vY_W$ is also independent of $\vM_{S \setminus I_C}$ and $\vK_{S \setminus I_C}$. Hence, by~\eqref{pf-equ11} and the independence of $(\vM_{I_C}, \vK_{I_C})$ and $(\vM_{S \setminus I_C}, \vK_{S \setminus I_C})$, we obtain that $\big(\sum_{i\in I_C}\vM_i, \vY_W\big)$ and $(\vM_{S \setminus I_C}, \vK_{S \setminus I_C})$ are independent, which immediately implies that
\begin{align}\label{pf-equ2}
H\bigg(\sum_{i\in I_C}\vM_i, \vY_W\bigg)=H\bigg(\sum_{i\in I_C}\vM_i, \vY_W \Big| \vM_{S\setminus I_C}, \vK_{S\setminus I_C}  \bigg).
\end{align}
Continuing from~\eqref{pf-equ2}, we consider
\begin{align}
&H\bigg(\sum_{i\in I_C}\vM_i, \vY_W \Big| \vM_{S\setminus I_C}, \vK_{S\setminus I_C}  \bigg)\nonumber \\
&=H\bigg(\sum_{i\in I_C}\vM_i \Big| \vY_W, \vM_{S\setminus I_C}, \vK_{S\setminus I_C}  \bigg)
 +H\big(\vY_W \big| \vM_{S\setminus I_C}, \vK_{S\setminus I_C}  \big)\nonumber \\
&=H\bigg(\sum_{i\in I_C}\vM_i \Big| \vY_W, \vM_{S\setminus I_C}, \vK_{S\setminus I_C}  \bigg)
 +H\big(\vY_W  \big),\label{pf-equ3}
\end{align}
where the equality~\eqref{pf-equ3} again follows from the independence of $\vY_W$ and $(\vM_{S\setminus I_C}, \vK_{S\setminus I_C})$. Now, we combine the equalities~\eqref{pf-equ1}, \eqref{pf-equ2} and~\eqref{pf-equ3} to obtain
\begin{align}\label{pf-equ4}
H\bigg(\sum_{i\in I_C}\vM_i\bigg)=H\bigg(\sum_{i\in I_C}\vM_i \Big| \vY_W, \vM_{S\setminus I_C}, \vK_{S\setminus I_C}  \bigg).
\end{align}

Next, we will prove the equality
\begin{align}\label{pf-equ5}
H \bigg( \sum_{i\in I_C}\vM_i \Big| \vY_{C},\vM_{S\setminus I_C},\vK_{S\setminus I_C} \bigg)=0.
\end{align}
We first consider the edge subset
$C' \triangleq \bigcup_{i\in {S\setminus I_C}} \eout(\sigma_i)$. Then,
$D_{C'}=I_{C'}=S\setminus I_C$. So, we see that $\vY_{C'}$ depends only on $\vM_{S\setminus I_C}$ and $\vK_{S\setminus I_C}$, and similar to~\eqref{pf-equ11}, we write
\begin{align*}
\vY_{C'}=\hg_{C'}(\vM_S,\vK_S)=\hg\,'_{C'}(\vM_{S\setminus I_C},\vK_{S\setminus I_C}).
\end{align*}
By the above discussion, we have
\begin{align}\label{pf-equ10}
\begin{split}
H\bigg(\sum_{i\in I_C}\vM_i \Big| \vY_{C},\vM_{S\setminus I_C},\vK_{S\setminus I_C}\bigg)
&=H\bigg(\sum_{i\in I_C}\vM_i \Big| \vY_{C},\vY_{C'},\vM_{S\setminus I_C},\vK_{S\setminus I_C}\bigg)\\
&=H\bigg(\sum_{i\in I_C}\vM_i \Big| \vY_{\hC},\vM_{S\setminus I_C},\vK_{S\setminus I_C} \bigg),
\end{split}
\end{align}
where $\hC = C \cup C'$. Clearly, $\hC$ is a global cut set, namely that $\hC$ separates $\rho$ from all the source nodes in $S$. Together with the acyclicity of the graph $\mG$, we see that $\vY_{\ein(\rho)}=\hg_{\ein(\rho)}(\vM_S,\vK_S)$ is a function of $\vY_{\hC}=\hg_{\hC}(\vM_S,\vK_S)$. Together with the admissibility of the secure network code $\hmbC$, the target function~$f$ must be computable with zero error on the global cut set $\hC$, i.e.,
\begin{align}\label{pf-equ7}
0=H\big(f(\vM_S)|\vY_{\hC}\big)=H\bigg(\sum_{i\in S}\vM_i \Big| \vY_{\hC}\bigg),
\end{align}
because otherwise $f$ cannot be computed at $\rho$ with zero error.
Now, we continue from \eqref{pf-equ10} to obtain
\begin{align}
& H\bigg(\sum_{i\in I_C}\vM_i \Big| \vY_{C},\vM_{S\setminus I_C},\vK_{S\setminus I_C}\bigg) = H\bigg(\sum_{i\in I_C}\vM_i \Big| \vY_{\hC},\vM_{S\setminus I_C},\vK_{S\setminus I_C}\bigg) \nonumber\\
& = H\bigg(\sum_{i\in I_C}\vM_i \Big| \vY_{\hC}, \sum_{i\in S}\vM_i, \vM_{S\setminus I_C},\vK_{S\setminus I_C} \bigg)\label{eq_Omin_7}\\
& =0,\label{eq_Omin_7-1}
\end{align}
where the equality~\eqref{eq_Omin_7} follows from \eqref{pf-equ7} and the equality~\eqref{eq_Omin_7-1} follows from
\begin{align*}
 H\bigg(\sum_{i\in I_C}\vM_i \Big|\sum_{i\in S}\vM_i, \vM_{S\setminus I_C} \bigg)=0.
\end{align*}
We thus have proved the equality \eqref{pf-equ5}.

Next, we combine \eqref{equ1}, \eqref{pf-equ4} and \eqref{pf-equ5} to obtain
\begin{align}
\ell\cdot \log q & =  H\bigg(\sum_{i\in I_C}\vM_i \bigg) = H\bigg(\sum_{i\in I_C}\vM_i \Big| \vY_W, \vM_{S\setminus I_C},\vK_{S\setminus I_C}\bigg)\nonumber\\
& =  H\bigg(\sum_{i\in I_C}\vM_i \Big| \vY_W, \vM_{S\setminus I_C}, \vK_{S\setminus I_C} \bigg)-H\bigg(\sum_{i\in I_C}\vM_i \Big| \vY_{C},\vM_{S\setminus I_C},\vK_{S\setminus I_C}\bigg)\nonumber\\
& = I\bigg(\sum_{i\in I_C}\vM_i;\vY_{C \setminus W} \Big| \vY_{W},\vM_{S\setminus I_C},\vK_{S\setminus I_C} \bigg) \nonumber \\
& \leq H\big(\vY_{C \setminus W} \big| \vY_{W},\vM_{S\setminus I_C},\vK_{S\setminus I_C} \big) \nonumber \\
& \leq H(\vY_{C \setminus W}) \leq \big| C\setminus W \big| \cdot \log q^n \nonumber \\
&=  n\cdot \big(|C|-|W| \big)\cdot \log q, \nonumber %\label{eq_Omin_8}
\end{align}
i.e.,
\begin{align}\label{pf-equ8}
\frac{\ell }{n}\leq |C|-|W|.
\end{align}
Since the inequality~\eqref{pf-equ8} holds for all the wiretap sets $W \subseteq C$ that satisfies $0\leq |W| \leq r$ and $D_W\subseteq I_C$, we obtain that
\begin{align}\label{pf-equ9}
\frac{\ell }{n}\leq \min_{W \in \mW_r: \atop W \subseteq C \text{ and } D_W \subseteq I_C} \big( |C|-|W| \big).
\end{align}
By considering all the cut sets $C\in \Lambda(\mN)$, we further obtain
\begin{align*}
\frac{\ell }{n} \leq %& \leq \min_{C\in \Lambda(\mN)} \min_{\text{all } W \subseteq C \text{ with }\atop 0\leq |W| \leq r \text{ and } D_W\subseteq I_C} |C|-|W|\\
%& = \min_{ W \in \mW_r } \min_{\text{all } C \in\Lambda(\mN) \text{ with } \atop C\supseteq W \text{ and } I_C \supseteq D_W } |C|-|W|\\
%&=
\min_{(W,C)\in \mW_r \times \Lambda(\mN):\atop  W\subseteq C \text{ and }  D_W \subseteq  I_C}  \big( |C|-|W| \big).
\end{align*}
Finally, since the above upper bound is valid for all the admissible secure network codes for the model $(\mN, f, r)$, we have proved \eqref{equ_upper_bound}, and hence the theorem is proved.
\end{IEEEproof}

\medskip

%Next, we present an upper bound and a lower bound on the upper bound~\eqref{equ_upper_bound} obtained in Theorem~\ref{thm_upper_capa} on the secure computing capacity $\hmC(\mN,f,r)$.

\begin{remark}\label{remk-0}
The upper bound on $\hmC(\mN,f,r)$ in Theorem~\ref{thm_upper_capa} is always nonnegative because for any $(W,C)$ satisfying the condition in the minimum, we have $W\subseteq C$, and so $|C|\geq|W|$.
\end{remark}

\begin{cor}\label{cor_up-b_low-b}
Let $C_{\min}=\min_{1\leq i \leq s}\mincut(\sigma_i,\rho)$. The upper bound on the secure computing capacity $\hmC(\mN,f,r)$ obtained in Theorem~\ref{thm_upper_capa} is bounded by
\begin{align}\label{ineq_cor_up-b_low-b}
C_{\min}-r \leq \min_{(W,C)\in \mW_r \times \Lambda(\mN):\atop  W\subseteq C \text{ and }  D_W \subseteq  I_C} \big( |C|-|W| \big)
\leq C_{\min}.
\end{align}
\end{cor}
\begin{IEEEproof}
We first prove the lower bound in \eqref{ineq_cor_up-b_low-b}. We consider
\begin{align}
&\min_{(W,C)\in \mW_r \times \Lambda(\mN):\atop  W\subseteq C \text{ and }  D_W \subseteq  I_C} \big( |C|-|W| \big) \nonumber\\
&= \min_{C \in\Lambda(\mN) } \min_{ W \in \mW_r: \atop W \subseteq C \text{ and } D_W \subseteq I_C } \big(|C|-|W|\big) \nonumber\\
&= \min_{ C \in\Lambda(\mN) } \Big( |C|-\max_{ W \in \mW_r:  \atop W \subseteq C \text{ and } D_W \subseteq I_C} |W| \Big) \nonumber\\
&\geq \min_{ C \in\Lambda(\mN) }|C|-r \label{ineq1-pf-ineq_cor_up-b_low-b}\\
&=C_{\min}-r,\label{ineq1-1-pf-ineq_cor_up-b_low-b}
\end{align}
where the inequality~\eqref{ineq1-pf-ineq_cor_up-b_low-b} follows from $|W|\leq r$, $\forall~W \in \mW_r$; and the equality~\eqref{ineq1-1-pf-ineq_cor_up-b_low-b} follows from the observation that
\begin{align}\label{ineq2-pf-ineq_cor_up-b_low-b}
C_{\min} = \min\big\{|C|:~C\in \Lambda(\mN)\big\},
\end{align}
which is explained as follows. First, we have
\begin{align}\label{ineq2-1-pf-ineq_cor_up-b_low-b}
C_{\min} \geq \min\big\{|C|:~C\in \Lambda(\mN)\big\},
\end{align}
because any minimum cut separating $\rho$ from a source node $\sigma_i$ is a cut set in $\Lambda(\mN)$.
On the other hand, consider a cut set
\begin{align*}
C^*\in \arg\min_{C}\big\{|C|:~C\in \Lambda(\mN) \big\}.
\end{align*}
Clearly, $C^*$ is a cut separating $\rho$ from a source node $\sigma$ in $I_{C^*}$. This implies
\begin{align*}
|C^*|\geq \mincut(\sigma,\rho)\geq C_{\min}.
\end{align*}
Together with~\eqref{ineq2-1-pf-ineq_cor_up-b_low-b}, we have proved \eqref{ineq2-pf-ineq_cor_up-b_low-b}.
%Otherwise, any minimum cut separating $\rho$ from $\sigma$, say $\hC$, satisfies $|\hC|<|C^*|$ and $\hC\in \Lambda(\mN)$, which contradicts the selection of $C^*$ in~\eqref{ineq2-1-pf-ineq_cor_up-b_low-b}.

Next, we prove the upper bound in \eqref{ineq_cor_up-b_low-b}. Rewrite the upper bound~\eqref{equ_upper_bound} in Theorem~\ref{thm_upper_capa} as
\begin{align}\label{ineq3-pf-ineq_cor_up-b_low-b}
\min_{(W,C)\in \mW_r \times \Lambda(\mN):\atop  W\subseteq C \text{ and }  D_W \subseteq  I_C} \big(|C|-|W|\big)
=\min_{ W \in \mW_r } \min_{C \in\Lambda(\mN):\atop  C\supseteq W \text{ and } I_C \supseteq D_W } \big(|C|-|W|\big).
\end{align}
Consider the empty set $\emptyset$, a particular wiretap set in $\mW_r$. Clearly, each cut set $C \in\Lambda(\mN)$ satisfies $\emptyset \subseteq C$ and ($\emptyset=$) $D_\emptyset \subseteq I_C$. By~\eqref{ineq3-pf-ineq_cor_up-b_low-b}, we immediately obtain that
\begin{align*}
\min_{ W \in \mW_r } \min_{C \in\Lambda(\mN):\atop  C\supseteq W \text{ and } I_C \supseteq D_W } \big(|C|-|W|\big) \leq \min_{ C \in\Lambda(\mN) }|C|-|\emptyset| = C_{\min}.
\end{align*}

Combining~\eqref{ineq1-pf-ineq_cor_up-b_low-b} and~\eqref{ineq3-pf-ineq_cor_up-b_low-b}, the corollary is thus proved.
\end{IEEEproof}

\medskip

Next, we give two corollaries, which are two cases such that the secure computing capacities are exactly characterized by the upper bound in Theorem~\ref{thm_upper_capa}.

\begin{cor}\label{cor-capacity=0}
Consider the model $(\mN, f, r)$, where the target function $f$ is the algebraic sum over a finite field $\Fq$. If the security level $r$ satisfies
\begin{align}\label{equ_cor_capacity=0}
%r\geq \min\big\{|C|:~C\in \Lambda(\mN) \text{ with } D_C=I_C \big\},
r\geq \overline{C}_{\min} \triangleq \min\big\{|C|:~C \in\Lambda(\mN) \text{ and } D_C=I_C  \big\},
\end{align}
then $$\hmC(\mN,f,r)=0.$$
\end{cor}
\begin{IEEEproof}
Let $W$ be an edge subset in the set
\begin{align}\label{equ1-cor}
\arg\min_{C}\big\{|C|:~ C \in\Lambda(\mN) \text{ and } D_C=I_C \big\}.
\end{align}
Since $I_W \neq \emptyset$, we readily see that $W\in \Lambda(\mN)$. On the other hand, by~\eqref{equ_cor_capacity=0} we have $|W|\leq r$ and so $W$ is a wiretap set in $\mW_r$. Together with $D_W=I_W$ from \eqref{equ1-cor}, $(W,W)$ is a pair in $\mW_r \times \Lambda(\mN)$ satisfying $W\subseteq W$ and $D_W \subseteq  I_W$. Then, the upper bound~\eqref{equ_upper_bound} obtained in Theorem~\ref{thm_upper_capa} is $0$, which, together with $\hmC(\mN,f,r)\geq0$, implies $\hmC(\mN,f,r)=0$.
\end{IEEEproof}

A straightforward consequence of Corollary~\ref{cor-capacity=0} is that $\hmC(\mN,f,r)=0$ if $r\geq \min_{1\leq i \leq s} \big|\eout(\sigma_i)\big|$, because $\eout(\sigma_i)$ satisfies $D_{\eout(\sigma_i)}=I_{\eout(\sigma_i)}\neq \emptyset$ for all $i$. Next, we consider the case of security level $r=0$. By Theorem~\ref{thm_upper_capa}, we immediately obtain that
\begin{align}\label{equ13}
   \hmC(\mN,f,0) & \leq
   \min_{(W,C)\in \mW_0 \times \Lambda(\mN):\atop  W\subseteq C \text{ \rm  and }  D_W \subseteq  I_C} \big(|C|-|W|\big) = \min_{ C\in \Lambda(\mN)}|C| =C_{\min}.
\end{align}
In fact, $C_{\min}$ is the computing capacity for the algebraic sum over an arbitrary network without security consideration (cf.\cite{HTYGuang-TIT18,Guang_NFC_TIT19}). Thus, the equation~\eqref{equ13} shows that the upper bound~\eqref{equ_upper_bound} thus obtained in Theorem~\ref{thm_upper_capa} degenerates to the computing capacity, denoted by $\mC(\mN,f)$, for the algebraic sum $f$ over a network $\mN$, which is stated in the following corollary.

\begin{cor}\label{cor4}
When the security level $r=0$, %the upper bound obtained in Theorem~\ref{thm_upper_capa} on the secure computing capacity $\hmC(\mN,f,r)$ degenerates to $C_{\min}$, i.e.,
\begin{align*}
   \hmC(\mN,f,0) = \mC(\mN,f) =C_{\min}.
\end{align*}
\end{cor}

However, we claim that the security level $r=0$ is not necessary for our upper bound to be equal to~$C_{\min}$. In other words, the secure computing capacity may still achieve $C_{\min}$ even though the security level $r$ is strictly larger than $0$. We illustrate this point by Example~\ref{exm_dis} below, which reveals the surprising fact that for some models of secure network function computation, there is no penalty on the secure computing capacity compared with the computing capacity without security consideration.

\begin{example}\label{exm_dis}

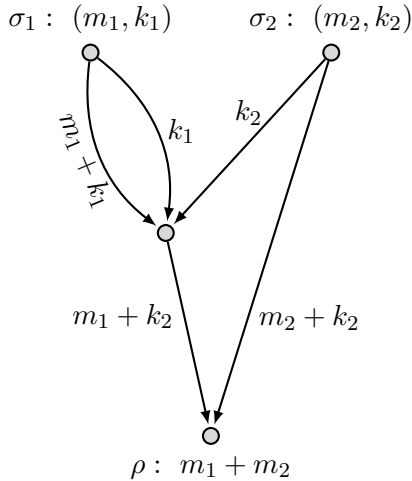
\begin{figure}[!t]
\centering
{
\begin{tikzpicture}
%[->,>=latex,node distance=2cm]
[->,>=latex,shorten >=1pt,auto,node distance=3cm, thick]
    \node[vertex]         (4)[label=below:$\rho:~m_1+m_2$]   {};
    \node[vertex]         (3)[above of=4, xshift=-6mm, yshift=-3mm]   {};
    \node[vertex]         (1)[above of=4, xshift=-16mm, yshift=21mm,label=above:{$\sigma_1:~(m_1,k_1)$}]   {};
    \node[vertex]         (2)[above of=4, xshift=16mm, yshift=21mm,label=above:{$\sigma_2:~(m_2,k_2)$}]   {};
\draw
(1) edge[bend left]           node[auto,swap,right=0mm] {$k_1$} (3)
    %edge                     node {} (3)
    edge[bend right]          node[auto,sloped, pos=0.5,below=-1mm] {$m_1+k_1$} (3)
(2) edge                      node[auto,swap, above=1mm] {$k_2$} (3)
    edge                      node[auto, below, pos=0.7, right] {$m_2+k_2$} (4)
    %edge[bend left]          node[auto,sloped, below] {\footnotesize$m_2+k_2$} (4)
(3) edge                      node[auto,swap, align=center, pos=0.4, left] {$m_1+k_2$}(4);
\end{tikzpicture}
}
\caption{An admissible $(1,1)$ secure network code for the secure model $(\mN_1,f,1)$.}\label{fig_N1}
\end{figure}

We consider securely computing the algebraic sum $f(m_1,m_2)=m_1+m_2$ on $\mathbb{F}_2$ over the network $\mN_1$ depicted in Fig.~\ref{fig_N1} with the security level $r=1$. For this secure model $(\mN_1,f,1)$, by Corollary~\ref{cor_up-b_low-b} we have
$$\hmC(\mN_1,f,1) \leq C_{\min}=1.$$
On the other hand, we present an admissible $(1,1)$ secure network code in Fig.~\ref{fig_N1}, of which the secure computing rate achieves the upper bound $C_{\min}=1$. Here, $k_i\in \mK_i=\mathbb{F}_2$ is an arbitrary output of the random key $\vK_i$ for the source node $\sigma_i$, $i=1,2$.
\end{example}

%\begin{fullversion}

\subsection{An Efficient Approach for Computing the Upper Bound}

In this subsection, we will provide an efficient approach for computing the upper bound \eqref{equ_upper_bound} in Theorem~\ref{thm_upper_capa}.
We consider a wiretap set $W \in \mW_r$ and define
\begin{align}\label{hat_C_W}
  \Omega(W)=\min_{C \in \Lambda(\mN): \atop W \subseteq C \text{ and } D_W \subseteq I_C} \big(|C|-|W| \big)=\min_{C \in \Lambda(\mN): \atop W \subseteq C \text{ and } D_W \subseteq I_C}|C\setminus W|.
\end{align}
Immediately, the upper bound in Theorem~\ref{thm_upper_capa} (cf.~\eqref{equ_upper_bound}) can be rewritten as
\begin{align}\label{UpperBound-v1}
  \hmC(\mN,f,r) \leq \min_{W \in \mW_r}\Omega(W).
\end{align}

%Before discussing further, we introduce some graph-theoretic concepts.
We first extend the definition of a cut separating a subset of nodes $V$ from another subset of nodes $U$ in the graph $\mG$ to the definition of a cut separating a subset of edges $W$ from a subset of nodes $U$ as follows. We subdivide each edge $e\in W$ by creating a node $v_e$ and splitting $e$ into two edges $e^1$ and $e^2$ such that $\tail(e^1)=\tail(e)$, $\head(e^2)=\head(e)$, and $\head(e^1)=\tail(e^2)=v_e$. Let $V_W=\big\{v_e: e\in W \big\}$. Then a {\em cut} separating the edge subset $W$ from the set of nodes $U$ is defined by a cut separating $V_W$ from $U$, where, whenever $e^1$ or $e^2$ appears in the cut, replace it by~$e$. By definition, we note that $W$ itself is a cut separating $W$ from $U$. Similarly, the minimum cut capacity separating $V_W$ from $U$ is defined as the \textit{minimum cut capacity} separating $W$ from $U$. Also, a cut separating $W$ from $U$ achieving this minimum cut capacity is called a \textit{minimum cut} separating $W$ from $U$. Furthermore, we say that a minimum cut separating $W$ from $U$ is {\em primary} if it separates from $U$ all the minimum cuts that separate $W$ from $U$. In particular, for a node $v\in \mV \setminus U$, a primary minimum cut separating $\ein(v)$ from $U$ will simply be referred to as a primary minimum cut separating $v$ from $U$. Similarly, we say that an edge subset $W \subseteq \mE$ is {\em primary} if $W$ is the primary minimum cut separating $W$ itself from $D_W$ in the graph $\mG$.
The concept of primary minimum cut was introduced by Guang and Yeung~\cite{GY-SNC-Reduction}, where its existence and uniqueness were proved. We now use a graph in \cite{GY-LNEC_revisted} (Fig.~2) as an example to illustrate the above graph-theoretic concepts.

\begin{example}
\begin{figure}[t]
\tikzstyle{vertex}=[draw,circle,fill=gray!30,minimum size=6pt, inner sep=0pt]
\tikzstyle{vertex1}=[draw,circle,fill=gray!80,minimum size=6pt, inner sep=0pt]
\centering
\begin{minipage}[b]{0.5\textwidth}
\centering
{
 \begin{tikzpicture}[x=0.6cm]
    \draw (0,0) node[vertex]     (s){};
    \draw (-2,-1.3) node[vertex] (1)[label=left:$u_1$]  {};
    \draw ( 2,-1.3) node[vertex] (2)[label=right:$u_2$]  {};
    \draw ( 0,-2.6) node[vertex] (3)  {};
    \draw ( 0,-4) node[vertex] (6)  {};
    \draw (-2,-6) node[vertex] (4)  {};
    \draw ( 1.5,-5.5) node[vertex] (5)  {};

    \draw[->,>=latex] (s) -- (1) node[midway, left=-0.2mm] {$e_1$};
    \draw[->,>=latex] (s) -- (2) node[midway, auto, right=-0.2mm] {$e_2$};
    %\draw[->,>=latex] (1) -- (4) node[midway, auto,swap, left=-1mm] {$e_5$};
    \draw[->,>=latex] (1) -- (3) node[midway, auto,left=-0.5mm] {$e_3$};
    \draw[->,>=latex] (2) -- (3) node[midway, auto,right=-0.5mm] {$e_4$};
    %\draw[->,>=latex] (2) -- (5) node[midway, auto, right=-1mm] {$e_8$};
    \draw[->,>=latex] (3) -- (6) node[midway, auto, left=-1mm] {$e_5$};
    \draw[->,>=latex] (6) -- (4) node[midway, auto, left=-0.5mm] {$e_{6}$};
    \draw[->,>=latex] (6) -- (5) node[midway, auto,swap, left=-0.5mm] {$e_{7}$};
    \draw[->,>=latex] (5) -- (4) node[midway, auto,swap, below=-0.2mm] {$e_{8}$};
    \end{tikzpicture}
}
\caption{The graph $\mG$.}
\label{fig-network1}
\end{minipage}%
\centering
\begin{minipage}[b]{0.5\textwidth}
\centering
 \begin{tikzpicture}[x=0.6cm]
    \draw (0,0) node[vertex]     (s)  {};
    \draw (-2,-1.3) node[vertex] (1)[label=left:$u_1$]  {};
    \draw ( 2,-1.3) node[vertex] (2)[label=right:$u_2$]  {};
    \draw ( 0,-2.6) node[vertex] (3)  {};
    \draw ( 0,-4) node[vertex] (6)  {};
    \draw ( 0.8,-4.8) node[vertex1] (8) [label={[xshift=-2mm, yshift=-5.5mm]$v_{e_7}$}] {};
   % \draw ( 0.8,-4.8) node[vertex1] (8) [label=right:$t_{e_7}$] {};
    \draw (-2,-6) node[vertex] (4)  {};
    \draw (-0.2,-5.75) node[vertex1] (9) [label={[xshift=0mm, yshift=-1mm]$v_{e_8}$}] {};
    \draw ( 1.5,-5.5) node[vertex] (5)  {};

    \draw[->,>=latex] (s) -- (1) node[midway, left=-0.2mm] {$e_1$};
    \draw[->,>=latex] (s) -- (2) node[midway, auto, right=-0.2mm] {$e_2$};
    %\draw[->,>=latex] (1) -- (4) node[midway, auto,swap, left=-1mm] {$e_5$};
    \draw[->,>=latex] (1) -- (3) node[midway, auto,left=-0.5mm] {$e_3$};
    \draw[->,>=latex] (2) -- (3) node[midway, auto,right=-0.5mm] {$e_4$};
    \draw[->,>=latex] (3) -- (6) node[midway, auto, left=-1mm] {$e_5$};

    \draw[->,>=latex] (6) -- (4) node[midway, auto, left=-0.5mm] {$e_{6}$};
    \draw[->,>=latex] (6) -- (8) node[pos=0.3, right=-0.5mm] {$e_{7}^1$};
    \draw[->,>=latex] (8) -- (5) node[pos=0.3, right=-0.5mm] {$e_{7}^2$};
    %\draw[->,>=latex] (5) -- (4) node[midway, auto,swap, below=-0.2mm] {};

    \draw[->,>=latex] (5) -- (9) node[pos=0.5, below=-1mm] {$e_8^1$};
    \draw[->,>=latex] (9) -- (4) node[pos=0.5, below=-1mm] {$e_8^2$};
    \end{tikzpicture}
\caption{The graph modification.}
\label{fig-network2}
\end{minipage}
\end{figure}

We consider a node subset $U=\{u_1,u_2\}$ and an edge subset $W=\{e_7,e_8\}$ in the graph $\mG$ depicted in Fig.~\ref{fig-network1}. For edge $e_7$, we first create a node $v_{e_7}$ and split $e_7$ into two edges $e_7^1$ and $e_7^2$ with $\tail(e_7^1)=\tail(e_7)$, $\head(e_7^2)=\head(e_7)$, and $\head(e_7^1)=\tail(e_7^2)=v_{e_7}$. The same subdivision operation is applied to edge $e_8$ as depicted in Fig.~\ref{fig-network2}. Let $V_W=\big\{v_{e_7}, v_{e_8}\big\}$. Now, in order to find a cut separating $W$ from $U$, it is equivalent to finding a cut separating $V_W$ from $U$. By definition, we see that the edge subset $\{e_3,e_4\}$ is a cut separating $W$ from $U$, because no path exists from $u_1$ or $u_2$ to $v_{e_7}$ or~$v_{e_8}$ upon removing $\{e_3,e_4\}$. Also, the edge subset $\{e_7^1\}$ is a cut separating $V_W$ from $U$. By definition, $e_7^1$ appears in the cut $\{e_7^1\}$ and $e_7\in W$, and thus $e_7^1$ is replaced by $e_7$ and so $\{e_7\}$ is a cut separating $W$ from $U$. We further see that $\{e_7\}$ is a minimum cut separating $W$ from $U$ that achieves the minimum cut capacity separating $W$ from $U$ whose value is $1$.

Furthermore, we see that $\{e_5\}$ is also a minimum cut separating $W$ from $U$, and $\{e_5\}$ and $\{e_7\}$ are all the minimum cuts separating $W$ from $U$. We further observe that $\{e_5\}$ separating from $U$ all the minimum cuts separating $W$ from $U$, i.e., two singleton subsets of edges $\{e_5\}$ and $\{e_7\}$. Therefore, by definition, $\{e_5\}$ is the primary minimum cut separating $W$ from $U$.
\end{example}

Next, we prove several lemmas that are instrumental in establishing Theorem~\ref{thm_upper_capa-2}.

\begin{lemma}\label{prop1}
Let $W$ be an edge subset and $W'$ be a minimum cut separating $W$ from $D_W$. Then,
\begin{align}\label{equ-prop}
D_{W'}=D_W.
\end{align}
\end{lemma}
\begin{IEEEproof}
First, since $W'$ is a cut separating $W$ from $D_W$, namely that any path from a source node in $D_W$ to an edge in $W$ must pass through an edge in $W'$, we obtain that $D_W\subseteq D_{W'}$. It thus suffices to prove that $D_{W'} \subseteq D_W$. This can be proved by contradiction as follows. Assume that there exists a source node $\sigma \in D_{W'}$ but $ \sigma \notin D_W$. %, i.e., $\sigma \rightarrow W'$ but $ \sigma \nrightarrow W$, where .
Let $e$ be an edge in $W'$ such that $\sigma \rightarrow e$. Then, $e \nrightarrow d$, $\forall~d\in W$, because otherwise there exists an edge $d\in W$ such that $e \rightarrow d$, which, together with $\sigma \rightarrow e$, implies $\sigma \rightarrow d$, i.e., $\sigma \in D_W$. This is a contradiction to the assumption that $\sigma \notin D_W$.

Now, we remove the edge $e$ from $W'$ to obtain a new edge subset $W'\setminus \{e\}$. By $e \nrightarrow d$ for all $d\in W$ as discussed above, we see that the edge subset $W'\setminus \{e\}$ also separates $W$ from $D_W$. This contradicts the assumption that $W'$ is a minimum cut separating $W$ from $D_W$. The lemma is proved.
\end{IEEEproof}

In the rest of the paper, for an edge subset $W \subseteq \mE$, we let $\hW$ be the primary minimum cut separating~$W$ from $D_W$. Then, $W$ is primary if and only if $W=\hW$.

\begin{lemma}\label{lem_hC_primary_cut}
 Consider a wiretap set $W \in \mW_r$. Let $W'$ be a minimum cut separating $W$ from $D_W$. Then,
  \begin{align*}
    \Omega(\hW) \leq \Omega(W') \leq \Omega(W).
  \end{align*}
\end{lemma}
\begin{IEEEproof}
Let $W \in \mW_r$ and let $C_W \in \Lambda(\mN)$ be an arbitrary but fixed subset of edges in
\begin{align}\label{equ1-effici-approach}
\arg\min_{C \in \Lambda(\mN)}\big\{|C|:~ W \subseteq C \text{ and } D_W \subseteq I_C \big\}.
\end{align}
By~\eqref{hat_C_W}, we readily see that
\begin{align*}
\Omega(W) = |C_W \setminus W|.
\end{align*}
Let $W'$ be a minimum cut separating $W$ from $D_W$, and let
\begin{align}\label{eq_C_hW}
  C' = W' \bigcup \big( C_W \setminus W \big).
\end{align}
We claim that $C'$ is a cut separating the sink node $\rho$ from $D_W$. To see this, assume the contrary that there exists a path $P$ from a source node $\sigma \in D_W$ to $\rho$ that does not pass through any edge in $C'$, or equivalently, the path $P$ does not pass through any edge in $W'$ or $C_W \setminus W$. Note that $C_W$ is a cut separating $\rho$ from $I_{C_W}$, and by \eqref{equ1-effici-approach} we have $D_W \subseteq I_{C_W}$. This immediately implies that $C_W$ is also a cut separating $\rho$ from $D_W$. So, the path $P$ must pass through an edge in $C_W$, or equivalently, in $W$ or $C_W \setminus W$. Together with $C_W \setminus W \subseteq C'$ and the assumption that the path $P$ does not pass through any edge in $C'$, the path $P$ must pass through an edge in $W$. Further, we see that the path $P$ does not pass through any edge in $W'$ either, because $P$ does not pass through any edge in $C'$ \big($=W' \cup \big( C_W \setminus W \big)$\big). This contradicts the fact that $W'$ is a cut separating $W$ from $D_W$. We thus have proved that $C'$ is a cut separating $\rho$ from $D_W$. This immediately implies $D_W \subseteq I_{C'}$. Together with $D_{W'}=D_W$ by Lemma~\ref{prop1}, we see that $C'$ is a cut set in $\Lambda(\mN)$ such that $W' \subseteq C'$ and $D_{W'} \subseteq I_{C'}$. Consequently, we obtain that
\begin{align*}
  \Omega(W') & \leq \big|C'\setminus W' \big|=\big| C' \big| - \big|W' \big| = \big|W' \bigcup \big( C_W \setminus W \big)\big| - \big|W' \big|\\
   & \leq \big| W' \big| + \big| C_W \setminus W \big| - \big| W' \big| = \big| C_W \setminus W \big|\\
   & = \Omega(W).
\end{align*}

Furthermore, by definition, the primary minimum cut $\hW$ separating $W$ from $D_W$ is the common minimum cut from $D_W$ separating all the minimum cuts that separate $W$ from $D_W$. Thus, in the above discussion, by replacing $W$ and $W'$ by $W'$ and $\hW$, respectively, we obtain that $\Omega(\hW)\leq \Omega(W')$. The lemma is thus proved.
\end{IEEEproof}

\medskip

For a set of edges $W\subseteq \mE$, let $\mG_W$ be the {\em residual graph} upon deleting the edges in $W$ from the graph~$\mG$.

\begin{lemma}\label{lem_C*W_size}
  For a wiretap set $W \in \mW_r$, let $C^*_W$ be the primary minimum cut separating $\rho$ from $D_W$ in the residual graph $\mG_W$. Then
  \begin{align*}
    \big|C^*_W\big| = \Omega(W).
  \end{align*}
\end{lemma}
\begin{IEEEproof}
Fix a wiretap set $W \in \mW_r$, and consider an arbitrary cut set $C \in \Lambda(\mN)$ satisfying $W \subseteq C$ and $D_W \subseteq I_C$. We readily see that $C$ is a cut in the graph $\mG$ separating $\rho$ from $D_W$. Next, we claim that $C \setminus W$ is a cut separating $\rho$ from $D_W$ in the residual graph $\mG_W$. To see this, assume the contrary that in $\mG_W$ there exists a path $P$ from a source node $\sigma \in D_W$ to $\rho$ that does not pass through any edge in $C \setminus W$. Then, $P$ is a path in $\mG$ from $\sigma$ to $\rho$ that does not pass through any edge in $C \setminus W$ or $W$, i.e., in $C$. This contradicts the fact mentioned above that $C$ is a cut in $\mG$ separating $\rho$ from $D_W$. Let $C^*_W$ be the primary minimum cut separating $\rho$ from $D_W$ in $\mG_W$. We immediately have
\begin{align}\label{eq-0_C*W_leq}
|C^*_W| \leq |C \setminus W| = |C|-|W|.
\end{align}
Considering all cut sets $C \in \Lambda(\mN)$ satisfying $W \subseteq C$ and $D_W \subseteq I_C$, by~\eqref{eq-0_C*W_leq}, we have
\begin{align}\label{eq_C*W_leq}
  |C^*_W| \leq \min_{C \in \Lambda(\mN): \atop W \subseteq C \text{ and } D_W \subseteq I_C} \big(|C|-|W|\big).
\end{align}

On the other hand, for a cut $C_W$ in $\mG_W$ that separates $\rho$ from $D_W$, we see that $C \triangleq C_W \bigcup W$ is a cut in $\mG$ that separates $\rho$ from $D_W$, because otherwise there exists a path in $\mG$ from a source node $\sigma \in D_W$ to $\rho$ that does not pass through any edge in $C_W$ or $W$, or equivalently, there exists a path in $\mG_W$ from $\sigma$ to $\rho$, which contradicts the fact that $C_W$ is a cut separating $\rho$ from $D_W$ in~$\mG_W$. In particular, we consider the primary minimum cut $C^*_W$
separating $\rho$ from $D_W$ in $\mG_W$, and let $C^* \triangleq C^*_W \bigcup W$. By the above discussion, $C^*$ is a cut set in $\Lambda(\mN)$ satisfying
$W \subseteq C^*$ and $D_W \subseteq I_{C^*}$. Thus, we have
\begin{align}\label{eq_C*W_geq}
  |C^*_W| = |C^*|-|W| \geq \min_{C \in \Lambda(\mN): \atop W \subseteq C \text{ and } D_W \subseteq I_C} \big(|C|-|W|\big),
\end{align}
where the first equality in \eqref{eq_C*W_geq} follows from $C^*_W \bigcap W=\emptyset$.
Then the lemma is proved by combining \eqref{eq_C*W_leq} and \eqref{eq_C*W_geq}.
%\begin{align*}
%  |C^*_W| =  \min_{C \in \Lambda(\mN): \atop W \subseteq C \text{ and } D_W \subseteq I_C} \big(|C|-|W|\big).
%\end{align*}
\end{IEEEproof}

\medskip

An important consequence of Lemmas~\ref{prop1} and~\ref{lem_hC_primary_cut} is given in the following lemma. This, together with Lemma~\ref{lem_C*W_size}, provides an equivalent expression of the upper bound in Theorem~\ref{thm_upper_capa} which can be used to compute the obtained upper bound efficiently.

Recall that an edge subset $W \subseteq \mE$ is primary if $W$ is the primary minimum cut separating $W$ itself from $D_W$ in the graph $\mG$, or equivalently,  $W=\hW$.

\begin{lemma}\label{lem8}
Let $\mW'_r\subseteq \mW_r$ be the collection of all the primary wiretap sets in $\mW_r$, i.e.,
\begin{align*}
\mW'_r = \big\{W \in \mW_r:~ W=\hW \big\}.
\end{align*}
Then,
\begin{align}\label{equ-lem8}
\min_{W \in \mW_r}\Omega(W)=\min_{W \in \mW'_r}\Omega(W).
\end{align}
\end{lemma}
\begin{IEEEproof}
By Lemma~\ref{lem_hC_primary_cut}, we immediately obtain
\begin{align}\label{equ1-lem8}
\min_{W \in \mW_r}\Omega(W) \geq \min_{W \in \mW_r}\Omega(\hW).
\end{align}
%where $\hW$ is the primary minimum cut separating $W$ from $D_W$ in the graph $\mG$.
For each $W\in \mW_r$, since $|\hW|\leq |W|$, we have $\hW\in \mW_r$, and so $\Omega(\hW) \geq \min_{W \in \mW_r}\Omega(W)$. Thus, $\min_{W \in \mW_r}\Omega(\hW) \geq \min_{W \in \mW_r}\Omega(W)$. Together with~\eqref{equ1-lem8}, we obtain
\begin{align}\label{equ2-lem8}
\min_{W \in \mW_r}\Omega(W) = \min_{W \in \mW_r}\Omega(\hW).
\end{align}

To prove \eqref{equ-lem8}, it suffices to show $\min_{W \in \mW_r}\Omega(\hW)=\min_{W \in \mW'_r}\Omega(W)$, which follows from the claim that for any wiretap set $W' \in \mW_r$, if there exists a $W \in \mW_r$ such that $W'=\hW$, then $W'=\widehat{W'}$. We now justify this claim. By Lemma~\ref{prop1}, we have $D_{W'}=D_{W}$. Then, $W'$ is the primary minimum cut separating~$W$ from $D_{W'}$. We now show that $W'$ is also the primary minimum cut separating itself from $D_{W'}$, i.e., $W'=\widehat{W'}$. If the contrary is true, we have $\widehat{W'}\neq W'$. Then $\widehat{W'}$ is also a minimum cut separating~$W$ from $D_{W}$ ($=D_{W'}$) but $W'$ does not separate $\widehat{W'}$ from $D_{W}$, a contradiction to the assumption that $W'=\hW$. The lemma is proved.
\end{IEEEproof}
%where
%\begin{align*}
%\widetilde{\mW}_r = \big\{ \hW \in \mW_r:~\text{$\exists$ $W \in \mW_r$ s.t. $\hW$ %is the primary minimum cut separating $W$ from $D_W$} \big\}.
%\end{align*}

%Next, we will prove $\widetilde{\mW}_r=\mW'_r$, which, together with \eqref{equ2-lem8}, immediately implies \eqref{equ-lem8}. First, we readily see that $\mW'_r \subseteq \widetilde{\mW}_r$ because by definition, each $W\in \mW'_r\subseteq \mW_r$ is the primary minimum cut separating $W$ itself from $D_{W}$. To see $\widetilde{\mW}_r \subseteq \mW'_r$, we consider a $\hW \in \widetilde{\mW}_r$ which is the primary minimum cut separating some wiretap set $W' \in \mW_r$ from $D_{W'}$. By Lemma~\ref{prop1}, we have $D_{W'}=D_{\hW}$ and thus we obtain that $\hW$ is the primary minimum cut separating $\hW$ from $D_{\hW}$, i.e., $\hW\in\mW'_r$. The lemma is proved.
%Thus, $\hW$ is the primary minimum cut separating $\hW$ itself from $D_{\hW}$ because otherwise there exists a minimum cut $W^*$ ($\neq \hW$) separating $\hW$ from $D_{\hW}$ and thus separating $W'$ from $D_{W'}$, a contradiction to the assumption that $\hW$ is the primary minimum cut separating $W'$ from $D_{W'}$. By definition, $\hW$ is primary. We thus have proved $\widetilde{\mW}_r\subseteq \mW'_r$. Together with $\mW'_r \subseteq \widetilde{\mW}_r$ as mentioned above, we have proved $\widetilde{\mW}_r=\mW'_r$.
%\end{IEEEproof}

Combining Lemma~\ref{lem8} with Lemma~\ref{lem_C*W_size}, we can rewrite the upper bound in Theorem~\ref{thm_upper_capa} as
\begin{align*}
  \hmC(\mN,f,r) \leq \min_{W \in \mW_r}\Omega(W)= \min_{W \in \mW'_r}\Omega(W) = \min_{W \in \mW'_r}|C^*_W|.
\end{align*}
This is formally stated in the following theorem.

\begin{thm}\label{thm_upper_capa-2}
Consider the model of secure network function computation $(\mN, f, r)$, where the target function $f$ is the algebraic sum over a finite field $\Fq$. The upper bound on the secure computing capacity $\hmC(\mN,f,r)$ in Theorem~\ref{thm_upper_capa} can be rewritten as
\begin{align*}
  \hmC(\mN,f,r) \leq \min_{W \in \mW'_r}|C^*_W|,
\end{align*}
where $C^*_W$ is the primary minimum cut separating $\rho$ from $D_W$ in the residual graph $\mG_W$.
\end{thm}

Theorem~\ref{thm_upper_capa-2} induces an efficient approach for computing the upper bound in Theorem~\ref{thm_upper_capa}. To be specific, in order to compute the upper bound in Theorem~\ref{thm_upper_capa}, it suffices to find the primary minimum cut separating~$\rho$ from $D_W$ in the residual graph $\mG_W$ for each primary wiretap set $W$ in $\mW'_r$. Guang and Yeung \cite{GY-SNC-Reduction} developed an efficient algorithm for finding the primary minimum cut separating a node from another one in a directed acyclic graph, and also proved that the computational complexity of this algorithm is linear time in the number of the edges in the graph. This algorithm can be straightforwardly extended to a linear-time algorithm for finding the primary minimum cut separating a node from a subset of nodes. % (cf.~\cite{GY-SNC-Reduction,GY-LNEC_revisted}).
So, for each primary minimum cut $W$ in $\mW'_r$, we can find the primary minimum cut $C^*_W$ separating $\rho$ from~$D_W$ in the residual graph $\mG_W$ in $\mO\big( |\mE| \big)$ time. Furthermore, by using the approach in \cite[Algorithm~1]{GY-LNEC_revisted}, we can determine $\mW'_r$, or equivalently, find all the primary wiretap sets in $\mW_r$ in $\mO\big( |\mW'_r|\cdot |\mE| \big)$ time (cf.~\cite[Algorithm~1]{GY-LNEC_revisted}). Combining these discussions, we can compute the upper bound in $\mO\big( |\mW'_r|\cdot |\mE| \big)$ time, which is linear in $|\mE|$.

%\end{fullversion}

\begin{DiscussionAchievality}

\section{Discussions on the Achievability}\label{result_dis}

For the upper bound in Theorem~\ref{thm_upper_capa}, both $D_{\min}-r \leq C_{\min}$ and $C_{\min} \leq D_{\min}-r$ are possible, and the upper bound on $\hmC(\mN,f,r)$ can be tight for both cases, as illustrated in the example below.

\begin{figure}[!t]
\begin{minipage}[b]{0.5\textwidth}
  \centering
{
\begin{tikzpicture}
[->,>=stealth',auto,node distance=1.5cm]
  \tikzstyle{every state}=[fill=none,draw=black,text=black,minimum size=4pt,inner sep=0pt]
    \node[state]         (7)[label=below:\footnotesize$\rho$:~$m_1+m_2$]   {};
    \node[state]         (5)[above left of=7, xshift=0mm, yshift=-3.5mm]   {};
    \node[state]         (6)[above right of=7, xshift=0mm, yshift=-3.5mm]   {};
    \node[state]         (4)[above of=7, xshift=0mm, yshift=0mm]   {};
    \node[state]         (3)[above of=4, xshift=0mm, yshift=0mm]   {};
    \node[state]         (2)[above right of=3, xshift=0mm, yshift=0mm),label=above:{\footnotesize$(m_2,k_2)$}]   {};
    \node[state]         (1)[above left of=3, xshift=0mm, yshift=0mm,label=above:{\footnotesize$(m_1,k_1)$}]   {};
\path
(1) edge          node[auto,swap,align=center,right=-6mm] {\footnotesize$m_1+k_1$} (5)
    edge          node[auto,swap, above=1mm] {\footnotesize$k_1$} (3)
(2) edge          node[auto,swap, align=center,right=-6mm] {\footnotesize$m_2+k_2$} (6)
    edge          node[auto,swap, above=1mm] {\footnotesize$k_2$} (3)
(3) edge          node[auto,swap,pos=0.7,right=-5mm] {\footnotesize$k_1+k_2$} (4)
(4) edge          node {} (5)
    edge          node {} (6)
(5) edge          node[auto,sloped, pos=0.4, below=-0.5mm] {\footnotesize$m_1+k_2$} (7)
(6) edge          node[auto,sloped, pos=0.4, below=-0.5mm] {\footnotesize$m_2+k_2$} (7);
\end{tikzpicture}
}
\caption{The network $\mN_1$}\label{fig_N1}
\end{minipage}
\hspace{0.3cm}
\begin{minipage}[b]{0.5\textwidth}
\centering
{
\begin{tikzpicture}
[->,>=stealth',auto,node distance=1.5cm]
  \tikzstyle{every state}=[fill=none,draw=black,text=black,minimum size=4pt,inner sep=0pt]
    \node[state]         (4)[label=below:\footnotesize$\rho$:~$m_1+m_2$]   {};
    \node[state]         (3)[above of=4, xshift=-6mm, yshift=0mm]   {};
    \node[state]         (1)[above of=4, xshift=-12mm, yshift=21mm,label=above:{\footnotesize$(m_1,k_1)$}]   {};
    \node[state]         (2)[above of=4, xshift=12mm, yshift=21mm,label=above:{\footnotesize$(m_2,k_2)$}]   {};
\path
(1) edge[bend left]           node[auto,swap,right=0mm] {\footnotesize$k_1$} (3)
    edge                      node {} (3)
    edge[bend right]          node[auto,sloped, pos=0.5,below=-1mm] {\footnotesize$m_1+k_1$} (3)
(2) edge                      node[auto,swap, above=1mm] {\footnotesize$k_2$} (3)
    edge                      node {} (4)
    edge[bend left]          node[auto,sloped, below] {\footnotesize$m_2+k_2$} (4)
(3) edge                      node[auto,swap, align=center ,left] {\footnotesize$m_1+k_2$}(4);
\end{tikzpicture}
}
\caption{The network $\mN_2$}\label{fig_N2}
\end{minipage}
\end{figure}

\begin{example}\label{exm_dis}
We consider securely computing the algebraic sum $f(x_1,x_2)=x_1+x_2$ on $\mathbb{F}_2$ over two networks $\mN_1$ and $\mN_2$ (depicted in Figs.~\ref{fig_N1} and~\ref{fig_N2}, respectively) with the security level $r=1$.

For the model $(\mN_1,f,r=1)$ (cf.~Fig.\,\ref{fig_N1}), we can see that
$D_{\min}-r=2-1=1$ and $C_{\min}=2$, i.e.,  $D_{\min}-r<C_{\min}$. By Theorem~\ref{thm_upper_capa}, we have
$\hmC(\mN_1,f,r=1) \leq D_{\min}-r=1$. On the other hand, the rate-1 secure network code in Fig.~\ref{fig_N1} achieves this upper bound.

For the model $(\mN_2,f,r=1)$ (cf.~Fig.\,\ref{fig_N2}), we have
$D_{\min}-r=3-1=2$ and $C_{\min}=1$. We obtain by Theorem~\ref{thm_upper_capa} that
$\hmC(\mN_1,f,r=1) \leq C_{\min}=1$.
Further, the rate-1 secure network code in Fig.~\ref{fig_N2} achieves this upper bound.
\end{example}

Moreover, for the upper bound obtained in Theorem~\ref{thm_upper_capa}, $\hmC(\mN,f,r) \leq C_{\min}$ holds for any security level. In fact, $C_{\min}$ is the capacity for computing an algebraic sum over a network without security consideration (cf.\cite{huang15,Guang_NFC_TIT19}), i.e., the case of $r=0$. So, the secure computing capacity $\hmC(\mN_2,f,r=1)=C_{\min}$ in Example~\ref{exm_dis} reveals the surprising fact that for some models of secure network function computation, there is no penalty on the secure computing capacity compared with the computing capacity without security consideration. On the other hand, when $r=0$,
\begin{align*}
\hmC(\mN,f,0) \leq \min\big\{ D_{\min},~C_{\min}\big\}=C_{\min}
\end{align*}
degenerates to the capacity for computing an algebraic sum over a network.

In addition, we also provide a lower bound on $\hmC(\mN,f,r)$ by designing a secure network coding scheme.

\begin{thm}\label{thm_lower_capa}
Let $\mN$ be a network and $f$ be the algebraic sum over a finite field $\Fq$. Then,
\begin{align*}
\hmC(\mN,f,r) \geq C_{\min}-r.
\end{align*}
\end{thm}

Here we give the idea of the code construction. We first consider computing $f$ over $\mN$ without security consideration and construct a rate-$C_{\min}$ network code $\mbC$. Such a code can be constructed by existing coding schemes, e.g., \cite{Koetter-CISS2004,Rai-Dey-TIT-2012}. Next, we will construct a rate-$(C_{\min}-r)$ secure network code $\hmbC$ for $(\mN,f,r)$ building on the code $\mbC$. For each source node $\sigma_i$, we partition the rate $C_{\min}$ into two sub-rates $C_{\min}-r$ and $r$, where the sub-rate $C_{\min}-r$ is used to transmit the source messages for computing the target function at the sink node and the sub-rate $r$ is used by the keys for guaranteeing the security level $r$. Toward this end, we take linear transformations as pre-encoding at the source nodes to mix the source messages and the keys. As such, we can obtain a rate-$(C_{\min}-r)$ secure network code $\hmbC$, which satisfies the computability condition \eqref{dc} and the security condition~\eqref{sc}.

Now, we compare the upper bound in Theorem~\ref{thm_upper_capa} and the lower bound in Theorem~\ref{thm_lower_capa}. Immediately, we see that the secure computing capacity is exactly characterized when $C_{\min}=D_{\min}$, as stated in the following.

\begin{cor}\label{cor_tight}
When $C_{\min}=D_{\min}$,
$$
\hmC(\mN,f,r)=C_{\min}-r.
$$
\end{cor}

In fact, the first secure model $(\mN_1,f,r=1)$ in Example~\ref{exm_dis} is an instance of the corollary. Further, the second secure model $(\mN_2,f,r=1)$ in Example~\ref{exm_dis} shows that the lower bound in general is not tight.

\end{DiscussionAchievality}

\section{Linear (Function-Computing) Secure Network Coding}\label{sec:LinearSNC}

In this section, we will discuss linear secure network coding for computing the algebraic sum. We will present a construction of admissible linear secure network codes for the secure model $(\mN,f,r)$, which immediately leads to a lower bound on the secure computing capacity $\hmC(\mN,f,r)$, where $f$ is defined in~\eqref{equ-alge_sum}.

\subsection{Linear Secure Network Coding for Computing the Algebraic Sum}

We will define linear (function-computing) secure network codes for the secure model $(\mN,f,r)$ in this subsection. Briefly speaking, a secure network code is said to be linear if its local encoding function for each edge is linear. Let $\ell$ and $n$ be two positive integers. We now define an $(\ell,n)$ {\em linear secure network code} over the finite field $\Fq$ for the secure model $(\mN, f, r)$.
%For each $i=1,2,\cdots,s$, we still use $M_i$ to denote the information source at the $i$th source node~$\sigma_i$ that is a random variable according to the uniform distribution on the finite field $\Fq$. Let $M_S=(M_{1},M_{2}, \cdots, M_s)$ and all the sources $M_i$, $1\leq i \leq s$ are mutually independent.
Each source node $\sigma_i$ sequentially generates $\ell$ i.i.d. random variables $M_{i,1}, M_{i,2}, \cdots, M_{i,\ell}$ with generic random variable $M_i$. We let $\vM_i=(M_{i,1},M_{i,2}, \cdots, M_{i,\ell})$ and $\vM_S=(\vM_1,\vM_2,\cdots,\vM_s)$. Furthermore, we continue to use $\vK_i$ to denote the random key available at the source node $\sigma_i$ which is now assumed to be distributed uniformly on the vector space $\Fq^{r_i}$, where $r_i$ is a nonnegative integer. Let $\vK_S=(\vK_1,\vK_2,\cdots,\vK_s)$.
%All the keys $\vK_i$ and the source messages $\vM_i$, $i=1,2,\cdots, s$ are mutually independent.

Now, for $i=1,2,\cdots,s$, we let $\vm_i \in \Fq^{\ell}$ and $\vk_i\in \Fq^{r_i}$ be arbitrary outputs of the source message~$\vM_i$ and the key $\vK_i$, respectively. Accordingly, let $\vm_S = (\vm_1,\vm_2,\cdots,\vm_s)$ and $\vk_S = (\vk_1,\vk_2,\cdots,\vk_s)$, i.e., $\vm_S$ and $\vk_S$ are two arbitrary outputs of $\vM_S$ and $\vK_S$, respectively. Further, we let $\vx_i=(\vm_i~\vk_i)$ for $i=1,2,\cdots,s$ and let $\vx_S=(\vx_1~\vx_2~ \cdots~\vx_s)$.

An $(\ell,n)$ linear secure network code $\hmbC=\big\{ \htheta_{e}:~e\in \mE;~  \hvarphi \big\}$ consists of a {\em linear} local encoding function~$\htheta_e$ for each edge $e$ and a decoding function $\hvarphi$ at the sink node $\rho$ as follows:
\begin{itemize}
  \item For each linear local encoding function $\htheta_e$,
  \begin{equation}\label{linear_snc_local}
\htheta_e:
\begin{cases}
\qquad \Fq^\ell \times \Fq^{r_i} \rightarrow \Fq^n , & \text{if } \tail(e) =\sigma_i \text{ for some } i,\\
\prod\limits_{d \in \ein(\tail(e))} \Fq^n \rightarrow \Fq^n , & \text{otherwise;}
\end{cases}
\end{equation}
or more precisely,
\begin{equation}\label{linear_snc_local-eqv-def}
\begin{cases}
\htheta_e(\vx_i) = \vx_i \cdot A_{i,e} =(\vm_i~~\vk_i)\cdot A_{i,e}, & \text{if } \tail(e) =\sigma_i \text{ for some } i,\medskip\\
\htheta_e\big(\,\vy_d:~d \in \ein(v)\big) = \sum\limits_{d \in \ein(v)} \vy_d \cdot A_{d,e}, & \text{otherwise;}
\end{cases}
\end{equation}
where $v\triangleq \tail(e)$ is a node in $\mV\setminus \{ S\cup \rho \}$, $A_{i,e}$ is an $\Fq$-valued matrix of size $(\ell+r_i) \times n$, $\vy_d\in \Fq^n$ is a row $n$-vector standing for the message transmitted on an edge $d$, and $A_{d,e}$ is an $\Fq$-valued matrix of size $n\times n$ called the {\em local encoding matrix} of the adjacent edge pair $(d,e)$ for the linear secure network code~$\hmbC$;
  \item The decoding function $\hvarphi$ at the sink node $\rho$ is a mapping from $\prod_{\ein(\rho)}\Fq^n$ to $\Fq^{\ell}$, which is used to compute at $\rho$ the algebraic sum $f$ with zero error.
\end{itemize}

With the linear encoding mechanism as described in~\eqref{linear_snc_local-eqv-def}, we readily see that $\vy_e$ for each $e\in \mE$ is a linear function of $\vm_S$ and $\vk_S$, i.e., each global encoding function $\hg_{e}$, induced by the linear local encoding functions $\htheta_e$, $e \in \mE$, is linear. Then, for each $e\in \mE$, there exists an $\Fq$-valued matrix $\mathbf{g}_{e}$ of size $(\ell s+\sum_{i=1}^{s}r_i) \times n$ such that
\begin{align*}
\hg_{e}(\vx_S) = \vx_S \cdot \mathbf{g}_{e} = \vy_e.
\end{align*}
In the rest of the paper we use $(\vm_S~\vk_S)$ to represent $\big((\vm_1~\vk_1)~(\vm_2~\vk_2)~ \cdots~(\vm_s~\vk_s)\big)$, i.e., $(\vm_S~\vk_S)=\vx_S$, for the convenience of discussion. This abuse of notation should cause no ambiguity and would greatly simplify the notation. Note that the $(\ell,n)$ linear secure network code as defined is sometimes referred to in the literature (e.g., \cite{Medard-Effros-Ho-Karger-Allerton03,vector-LNC-ref2,vector-LNC-ref3,vector-LNC-ref5}) as a ``vector-linear'' code when $n>1$.

We end this subsection with two remarks on the definition of linear secure network codes.

\begin{remark}\label{remk-1}
In the definition of linear secure network codes, the decoding function $\hvarphi$ is not necessarily linear. Nevertheless, it can be proved that for any admissible linear secure network code $\hmbC=\big\{ \htheta_{e}:~e\in \mE;~  \hvarphi \big\}$ for computing the algebraic sum, there always exists a linear decoding function $\hvarphi'$ such that the code $\big\{ \htheta_{e}:~e\in \mE;~  \hvarphi' \big\}$ is still admissible.
\end{remark}

\begin{remark}
If $r_i=0$ in~\eqref{linear_snc_local} for all $i=1,2,\cdots,s$, then the definition of linear secure network codes for the secure model $(\mN,f,r)$ degenerates to the definition of linear network codes for the model $(\mN,f)$.
\begin{details}
To be specific, an $(\ell,n)$ linear network code $\mbC=\big\{ \theta_{e}:~e\in \mE;~  \varphi \big\}$ consists of a linear local encoding function $\theta_e$ for each edge $e$ and a decoding function $\varphi$ at the sink node as follows:
\begin{itemize}
  \item For each linear local encoding function $\theta_e$,
  \begin{equation}\label{linear_snc_local-sss}
\theta_e:
\begin{cases}
\qquad \Fq^\ell \rightarrow \Fq^n , & \text{if } \tail(e) =\sigma_i \text{ for some } i,\\
\prod\limits_{d \in \ein(\tail(e))} \Fq^n \rightarrow \Fq^n , & \text{otherwise;}
\end{cases}
\end{equation}
  \item The decoding function $\varphi:~\prod_{\ein(\rho)}\Fq^n \rightarrow \Fq^{\ell}$  at the sink node $\rho$ in order to compute the algebraic sum $f$ with zero error.
\end{itemize}
\end{details}
We say a linear network code for $(\mN,f)$ is {\em admissible} if the target function $f$ can be computed at the sink node with zero error, i.e., only the computability condition is satisfied.
\end{remark}

\subsection{Lower Bound on the Secure Computing Capacity}

We consider the secure model $(\mN, f, r)$ with security level $0 \leq r \leq C_{\min}$, where we recall that $C_{\min}\triangleq \min_{1\leq i \leq s}\mincut(\sigma_i,\rho)$. For any nonnegative integer~$R$ with $r \leq R \leq C_{\min}$, we can construct an admissible $(R-r, 1)$ linear secure network code for the model $(\mN, f, r)$. This code construction immediately provides a lower bound on the secure computing capacity $\hmC(\mN, f, r)$, which is stated in the following theorem. The proof of the theorem, including the code construction and the verification of the computability and security conditions, is deferred to Sections~\ref{sec: LSNC-construction} and~\ref{sec: LSNC-verification}.

\begin{thm}\label{thm_lower_capa}
Consider the model of secure network function computation $(\mN, f, r)$, where the target function $f$ is the algebraic sum over a finite field $\Fq$ and the security level $r$ satisfies $0 \leq r \leq C_{\min}$. Then
  \begin{align*}
    \hmC(\mN, f, r) \geq C_{\min} - r.
  \end{align*}
\end{thm}

We recall from \eqref{ineq2-pf-ineq_cor_up-b_low-b} and \eqref{equ_cor_capacity=0} that
\begin{align}\label{C_min}
C_{\min} = \min\big\{|C|:~C\in \Lambda(\mN)\big\},
\end{align}
and
\begin{align}\label{overline_C_min}
\overline{C}_{\min} \triangleq  \min\big\{|C|:~C\in \Lambda(\mN) \text{ and } D_C=I_C \big\}.
\end{align}
Clearly, we have $C_{\min} \leq \overline{C}_{\min}$.
By applying the lower bound in Theorem~\ref{thm_lower_capa}, the secure computing capacity $\hmC(\mN,f,r)$ can be fully characterized when $C_{\min} =\overline{C}_{\min}$, i.e., for some source node $\sigma_i \in S$, there exists a minimum cut $C$ separating $\rho$ from $\sigma_i$ such that $|C|=C_{\min}$ and $D_C=I_C$. This is given in the next corollary.

\begin{cor}\label{cor10}
When $C_{\min} = \overline{C}_{\min}$,
\begin{equation*}
\hmC(\mN,f,r)=
\begin{cases}
C_{\min}-r, & \text{if } r< C_{\min},\medskip\\
0 , & \text{otherwise}.
\end{cases}
\end{equation*}
\end{cor}
\begin{IEEEproof}
First, by Corollary~\ref{cor-capacity=0} we have
\begin{align}\label{ineq-cor}
\hmC(\mN,f,r)=0, ~~\forall~ r\geq \overline{C}_{\min}.
\end{align}
When $C_{\min} = \overline{C}_{\min}$, it immediately follows from \eqref{ineq-cor} that $\hmC(\mN, f, r) = 0$ for $r \geq C_{\min}$.

Next, we consider the case of $r < C_{\min}$. As $C_{\min} = \overline{C}_{\min}$, in light of~\eqref{overline_C_min}, there exists a cut set $\hC \in \Lambda(\mN)$ such that $D_{\hC} = I_{\hC}$ and $|\hC| = C_{\min}$. Now, we let $\widehat{W}$ be an edge subset of $\hC$ with $|\widehat{W}| = r$. Immediately, we see that
$\widehat{W} \in \mW_r$ and $D_{\widehat{W}} \subseteq D_{\hC} = I_{\hC}$. By the upper bound \eqref{equ_upper_bound} in Theorem~\ref{thm_upper_capa}, we have
\begin{align*}
  \hmC(\mN, f, r) & \leq \min_{(W,C)\in \mW_r \times \Lambda(\mN):\atop  W\subseteq C \text{ \rm  and }  D_W \subseteq  I_C} \big(|C|-|W|\big) \\
   & \leq |\hC| - |\widehat{W}| = C_{\min} - r.
\end{align*}
Together with the lower bound $\hmC(\mN, f, r) \geq C_{\min} - r$ in Theorem~\ref{thm_lower_capa}, we have proved that
\begin{align*}
  \hmC(\mN, f, r) =  C_{\min} - r ~\text{ for }~ r < C_{\min}.
\end{align*}
The corollary is thus proved.
\end{IEEEproof}

\medskip

When $C_{\min} < \overline{C}_{\min}$, by Corollary~\ref{cor-capacity=0}, $\hmC(\mN, f, r)=0$ if $r \geq \overline{C}_{\min}$. When $C_{\min} < \overline{C}_{\min}$ and the security level $r$ satisfies $r \leq C_{\min}$, we give in the next corollary a sufficient condition in terms of the network topology for the tightness of the lower bound in Theorem~\ref{thm_lower_capa}. However, for $C_{\min} < r < \overline{C}_{\min}$, a nontrivial lower bound\footnote{Here, we regard $0$ as a trivial lower bound on $\hmC(\mN, f, r)$.} on $\hmC(\mN, f, r)$ or a code construction is yet to be obtained.

\begin{cor}
When $C_{\min} < \overline{C}_{\min}$,
 \begin{align*}
  \hmC(\mN,f,r)=C_{\min}-r,\quad \forall~ r \leq C_{\min},
 \end{align*}
provided that for some source node $\sigma_i\in S$ with $\mincut(\sigma_i,\rho)=C_{\min}$,  there exists a minimum cut $C$ separating $\rho$ from $\sigma_i$ such that there are $r$ edges in $C$ satisfying that for each such edge $e$,
$$\sigma \nrightarrow e,\quad \forall~\sigma\in S\setminus I_C.\footnotemark$$
\end{cor}
\footnotetext{Note that besides $\sigma_i$, other source nodes may also be contained in $I_C$.}

\begin{IEEEproof}
Let $\sigma_i$ be a source node in $S$ such that
\begin{enumerate}
  \item $\mincut(\sigma_i,\rho)=C_{\min}$, and
  \item  there exists a minimum cut $C$ separating $\rho$ from $\sigma_i$ such that there are $r$ edges in $C$ satisfying that for each edge $e$ of the $r$ edges, $\sigma \nrightarrow e$, $\forall~\sigma\in S\setminus I_C$.
\end{enumerate}
Let $W$ be any subset of $r$ edges in $C$ satisfying the above condition 2). Then, we have
\begin{align*}
W\subseteq C ~\text{ and }~ D_W\subseteq I_C.
\end{align*}
By $W\in\mW_r$ (since $|W|=r$) and the upper bound in Theorem~\ref{thm_upper_capa}, we thus obtain that
\begin{align*}
\hmC(\mN,f,r)\leq |C|-|W|=C_{\min}-r.
\end{align*}
Together with the lower bound in Theorem~\ref{thm_lower_capa}, we have proved that $\hmC(\mN,f,r)=C_{\min}-r$.
\end{IEEEproof}

\subsection{Code Construction}\label{sec: LSNC-construction}

Before presenting the code construction for the secure model $(\mN, f, r)$ with security level $0 \leq r \leq C_{\min}$, we first consider the model $(\mN, f)$, i.e., computing the algebraic sum $f$ over $\mN$ without any security constraint. For any nonnegative integer $R$ with $r \leq R \leq C_{\min}$, we can construct an admissible $(R, 1)$ linear network code for $(\mN, f)$ by ``reversing'' a rate-$R$ linear network code for the single-source multicast problem on the {\em reversed network} $\mN^\top \triangleq (\mG^\top,S,\rho)$, where the {\em reversed graph} $\mG^\top$ is obtained from $\mG$ by reversing the direction of every edge in $\mG$, and $\rho$ and $S$ are regarded as the single source node and the set of sink nodes, respectively. For the completeness of the paper, we include a discussion on the construction of such an admissible $(R, 1)$ linear network code for $(\mN, f)$ in Appendix~\ref{appendix} (also cf.~\cite{Koetter-CISS2004,Rai-Dey-TIT-2012}\footnote{In Appendix~\ref{appendix}, we use a global approach to describe the relation between a linear network code and its ``reversed'' code, which is different from the local approach in \cite{Koetter-CISS2004,Rai-Dey-TIT-2012}. Nevertheless, the essence of the two approaches are the same.}). In the following, we will construct an admissible $(R-r, 1)$ linear secure network code for the secure model $(\mN, f, r)$  based on an admissible $(R, 1)$ linear network code for $(\mN, f)$.

By the acyclicity of the graph $\mG$, we can fix a topological order ``$\prec$'' on the edges in $\mE$ that is consistent with the natural partial order of the edges. To facilitate our discussion, we further assume without loss of generality that this order ``$\prec$'' satisfies the following two conditions: \rmnum{1}) the output edges of the source nodes in $\bigcup_{i=1}^s\eout(\sigma_i)$ are prior to the other edges in $\mE \setminus \bigcup_{i=1}^s\eout(\sigma_i)$, i.e.,
\begin{align*}
d \prec e, \quad \forall~d \in \bigcup_{i=1}^s\eout(\sigma_i) \quad  \text{and} \quad  e \in \mE \setminus \bigcup_{i=1}^s\eout(\sigma_i);
\end{align*}
and \rmnum{2}) for any two edges $e_i\in \eout(\sigma_i)$ and $e_j\in \eout(\sigma_j)$,
\begin{align*}
e_i \prec e_j ~\text{ if }~ i < j.
\end{align*}
In the rest of the paper, we always use this order to index the coordinates of vectors and the rows/columns of matrices.

Let $R$ be a positive integer not larger than $C_{\min}$. Let $\vx_i \triangleq (x_{i, 1}, x_{i, 2}, \cdots, x_{i, R})\in \Fq^R$ be the vector of $R$ symbols sequentially generated by the source node $\sigma_i$ for $i=1, 2, \cdots, s$, and let $\vx_S = (\vx_1, \vx_2, \cdots, \vx_s)$. Further, let $\mbC = \big\{\theta_e: e \in \mE;~\varphi \big\}$ be an admissible $(R, 1)$ linear network code for computing the algebraic sum $f$ over the network $\mN$, where $\theta_e$ is the local encoding function of edge~$e$. By the linearity of $\mbC$, $\theta_e$ satisfies
\begin{align}\label{equ_R_1_lnc}
\begin{cases}
\theta_e(\vx_i)=\vx_i \cdot A_{i,e}, & \text{if } \tail(e) =\sigma_i \text{ for some } i,\medskip\\
\theta_e\big(\,y_d:~d \in \ein(v)\big) = \sum\limits_{d \in \ein(v)}a_{d,e} \cdot y_d, & \text{otherwise,}
\end{cases}
\end{align}
where $v = \tail(e)$, $A_{i,e}$ is a column $R$-vector in $\Fq^R$, $y_e \in \Fq$ is the message transmitted on an edge~$e$ for the source messages $\vx_S$, and $a_{d,e}\in \Fq$ is the local encoding coefficient of the adjacent edge pair $(d, e)$. We further let $g_e$, $e \in \mE$ be the linear global encoding functions induced by the linear local encoding functions $\theta_e$, $e \in \mE$, i.e.,
\begin{align*}
y_e = g_e(\vx_S),\qquad \forall~e\in \mE.
\end{align*}
By the linearity of the global encoding function $g_e$, there exists an $\Fq$-valued column $R s$-vector $\vg_e$ such that
\begin{align}\label{equ_linear-global-encoding-vec}
g_e(\vx_S) = \vx_S \cdot \vg_e.
\end{align}
We call $\vg_e$ the {\em global encoding vector} for the edge $e$. We further write
\begin{align}\label{ge-vec-form}
\vg_e=\begin{bmatrix}\vg_e^{\,(\sigma_1)} \\ \vg_e^{\,(\sigma_2)} \\ \vdots \\ \vg_e^{\,(\sigma_s)}\end{bmatrix},
\end{align}
where each $\vg_e^{\,(\sigma_i)}$ is a column $R$-vector for $i=1,2,\cdots, s$. Then, the equation~\eqref{equ_linear-global-encoding-vec} can be written as
\begin{align}
g_e(\vx_S) = \vx_S \cdot \vg_e = \sum_{i=1}^{s}\vx_i \cdot \vg_e^{\,(\sigma_i)}.
\end{align}

Next, we will elaborate on how to calculate the global encoding vectors from the linear local encoding functions. To facilitate our discussion, for each source node $\sigma_i$, we introduce $R$ imaginary edges $d'_{i,1}, d'_{i, 2}, \cdots, d'_{i, R}$ connecting to $\sigma_i$, and let $\ein(\sigma_i)=\big\{d'_{i,1}, d'_{i, 2}, \cdots, d'_{i, R} \big\}$. We further assume that the~$R$ source messages $x_{i, j}$, $1\leq j \leq R$ are transmitted to $\sigma_i$ through these $R$ imaginary edges. To be specific, without loss of generality, assume that the source message $x_{i, j}$ is transmitted to $\sigma_i$ through the $j$th imaginary edge $d'_{i, j}$ for $j=1,2,\cdots,R$, i.e.,
\begin{align*}
y_{d'_{i, j}} = x_{i, j}, \quad 1\leq j \leq R,
\end{align*}
where $y_{d'_{i, j}}$ stands for the symbol transmitted on $d'_{i, j}$. For each source node $\sigma_i$, we define an $R \times |\mE|$ matrix
\begin{align}\label{equ_A-sigma-i}
A_{\sigma_i}=
  \begin{bmatrix}
    A_{\sigma_i,e}:~e \in \mE
  \end{bmatrix},
\end{align}
where
\begin{align*}
  A_{\sigma_i, e}=
  \begin{cases}
    A_{i, e}, & \mbox{if $e \in \eout(\sigma_i)$;} \\
    \bzero,   & \mbox{otherwise;\footnotemark}
  \end{cases}
\end{align*}
\footnotetext{Here, $\bzero$ is an all-zero column $R$-vector. In the following, we always use $\bzero$ to denote an all-zero column vector whose dimension is clear from the context.}
(cf.~\eqref{equ_R_1_lnc} for $A_{i, e}$).
For each intermediate node $v \in \mV \setminus (S \cup \{\rho\})$, we define an $|\ein(v)| \times |\eout(v)|$ matrix
\begin{align}\label{equ_A-v}
 A_v = \big(a_{d, e} \big)_{d \in \ein(v),~e \in \eout(v)},
\end{align}
where $a_{d, e}$ is the local encoding coefficient of the adjacent edge pair $(d, e)$ (cf.~\eqref{equ_R_1_lnc}). Further, we extend the local encoding coefficient $a_{d, e}$ to the case of non-adjacent edge pairs $(d, e)$ by setting $a_{d, e}=0$ if $\head(d) \neq \tail(e)$.
In particular, we have $a_{e, e}=0$, $\forall~e \in \mE$. Now, we let
\begin{align}\label{equ_A}
A = \big( a_{d, e} \big)_{d \in \mE,\, e \in \mE}
\end{align}
be an $|\mE| \times |\mE|$ matrix in which all the entries are well defined. Then $A$ is upper triangular and all the diagonal elements are equal to zero.

Similar to linear network coding (cf.~\cite{Koetter-Medard-algebraic}), we have
\begin{align}\label{G_rho_sigma}
  \begin{bmatrix}
    \vg_{e}^{\,(\sigma_i)}:~e\in \mE
  \end{bmatrix}
    = A_{\sigma_i} \cdot \big( I-A \big)^{-1},
\end{align}
where $I$ is the $|\mE| \times |\mE|$ identity matrix. Together with \eqref{ge-vec-form}, we further have
\begin{align}\label{G_rho}
  \begin{bmatrix}
    \vg_{e}:~e\in \mE
  \end{bmatrix}
    =
  \begin{bmatrix}
    \vg_{e}^{\,(\sigma_1)}:~e\in \mE \\
    \vg_{e}^{\,(\sigma_2)}:~e\in \mE \\
    \vdots \\
    \vg_{e}^{\,(\sigma_s)}:~e\in \mE
  \end{bmatrix}
  =
  \begin{bmatrix}
    A_{\sigma_1}\cdot\big( I-A \big)^{-1} \\
    A_{\sigma_2}\cdot\big( I-A \big)^{-1}  \\
    \vdots \\
    A_{\sigma_s}\cdot\big( I-A \big)^{-1}
  \end{bmatrix}
=
  \begin{bmatrix}
    A_{\sigma_1} \\
    A_{\sigma_2} \\
    \vdots \\
    A_{\sigma_s}
  \end{bmatrix}
  \cdot
\big( I-A \big)^{-1}.
\end{align}
Immediately, for each edge $e \in \mE$, it follows from \eqref{G_rho_sigma} and \eqref{G_rho} that
\begin{align}\label{equ_global_en_vector-i}
\vg_{e}^{\,(\sigma_i)}
    =
    A_{\sigma_i}
    \cdot
    \big( I-A \big)^{-1}\cdot \vec{1}_e, \quad \forall~i=1, 2, \cdots, s,
\end{align}
and
\begin{align}\label{equ_global_en_vector}
\vg_{e} =
  \begin{bmatrix}
    A_{\sigma_1} \\
    A_{\sigma_2} \\
    \vdots \\
    A_{\sigma_s}
  \end{bmatrix}
  \cdot
\big( I-A \big)^{-1}
\cdot
\vec{1}_e,
\end{align}
where $\vec{1}_{e}$ is a column $|\mE|$-vector whose component indexed by the edge $e$ is equal to $1$ while all other components are equal to $0$.

Based on the admissible $(R, 1)$ linear network code $\mbC = \big\{ \theta_e: e \in \mE;~\varphi \big\}$ for the model $(\mN, f)$, by using a similar approach as the one in \cite{Cai-Yeung-SNC-IT},  we will construct an admissible $(R-r, 1)$ linear secure network code $\hmbC = \big\{ \htheta: e \in \mE;~\hvarphi \big\}$ for the secure model $(\mN, f, r)$ with any security level $r \leq R$.
First, we let $\vb_1, \vb_2, \cdots, \vb_{R}$ be $R$ linearly independent column vectors in $\Fq^R$ such that the first $R-r$ vectors $\vb_1, \vb_2, \cdots, \vb_{R-r}$ satisfy the condition that for all the wiretap sets $W \in  \mW_r$ and all $i = 1, 2, \cdots, s$,
\begin{align}\label{equ-sec-condition}
\big\langle  \vb_j:~ 1\leq j \leq R-r  \big\rangle \bigcap
\big\langle \vg_e^{\,(\sigma_i)}:~ e \in W  \big\rangle=\{\bzero\}.\footnotemark
\end{align}
\footnotetext{Here, we use $\langle L \rangle$ to denote the vector space spanned by the vectors in a set of vectors $L$.}The existence of such $R$ vectors will be discussed in the following Section~\ref{sec: LSNC-field_size}. Then, let
\begin{align*}
B = \left[ \vb_1 ~~ \vb_2 ~~ \cdots ~~ \vb_R \right],
\end{align*}
an $R \times R$ invertible matrix in $\Fq$, and
\begin{align}\label{equ-hat_B}
\hB = \big[ B~~B~~\cdots~~B  \big]_s^{\text{diagonal}}
\triangleq
\begin{bmatrix}
B & \bfzero & \cdots & \bfzero &\\
\bfzero & B & \cdots & \bfzero &\\
\cdots  & \cdots & \cdots & \cdots\\
\bfzero & \bfzero & \cdots & B\\
\end{bmatrix}_{R s \times R s},
\end{align}
an $s\times s$ block matrix in which each block in the diagonal is $B$ and all the other blocks are the $R \times R$ zero matrix.\footnote{Here, $\bfzero$ stands for an all-zero $R\times R$ matrix. In the following, we always use $\bfzero$ to denote an all-zero square matrix whose size is clear from the context.} Since the matrix $B$ is invertible, $\hB$ is also invertible and
\begin{align}\label{equ_hat_B_inverse}
\hB^{-1} = \big[ B^{-1}~~B^{-1}~~\cdots~~B^{-1}  \big]_s^{\text{diagonal}}.
\end{align}

We recall that at the source node $\sigma_i$, the information source $M_i$ is distributed uniformly on the finite field $\Fq$, $i=1,2,\cdots,s$. Let $\vM_i=(M_{i,1},M_{i,2}, \cdots, M_{i,R-r})$ be the vector of $R-r$ i.i.d. random variables generated by $\sigma_i$ with the generic random variable $M_i$, and $\vM_S=(\vM_1,\vM_2,\cdots,\vM_s)$. Furthermore, we let $K_i$ be a random variable that is distributed uniformly on $\Fq$. Let the key available at the source node~$\sigma_i$ be $\vK_i = (K_{i, 1}, K_{i, 2}, \cdots, K_{i, r})$, where $K_{i, 1}, K_{i, 2}, \cdots, K_{i, r}$ are $r$ i.i.d random variables with generic random variable $K_i$. Then, the key $\vK_i$ can be regarded as a random variable distributed uniformly on the vector space $\Fq^r$. Let $\vK_S=(\vK_1,\vK_2,\cdots,\vK_s)$. All the keys $\vK_i$ and the source messages $\vM_i$, $i=1,2,\cdots, s$ are mutually independent.
Let $\vm_i \in \Fq^{R-r}$ and $\vk_i\in \Fq^{r}$ be arbitrary outputs of the source message $\vM_i$ and the key $\vK_i$, respectively, for $i=1,2,\cdots,s$. Accordingly, let $\vm_S = (\vm_1,\vm_2,\cdots,\vm_s)$ and $\vk_S = (\vk_1,\vk_2,\cdots,\vk_s)$, and let $\vx_i=(\vm_i~\vk_i)\in \Fq^R$ and $\vx_S=(\vx_1~\vx_2~ \cdots~\vx_s)$.

By applying the linear network code $\mbC$ and the matrix $B$, we define the linear local encoding function~$\htheta_e$ for each edge $e \in \mE$ as follows:
\begin{align*}
\begin{cases}
\htheta_e\big(\vx_i\big) = \vx_i \cdot B^{-1} A_{i, e} = (\vm_i, \vk_i) \cdot B^{-1} A_{i, e}, & \text{if } \tail(e)=\sigma_i \text{ for some } i,~1\leq i \leq s,\medskip\\
\htheta_e\big(y_d,~d \in \ein(v)\big) = \theta_e\big(y_d,~d \in \ein(v)\big) = \sum_{d \in \ein(v)} a_{d, e} \cdot y_d, & \text{otherwise,}
\end{cases}
\end{align*}
where $v = \tail(e)$, an intermediate node in $\mV \setminus (S \cup \{\rho\})$, and  $y_d \in \Fq$ is the message transmitted on edge $d$ for the source message vector $\vm_S$ and the key vector $\vk_S$.
%%%%%%%%%%%%%%%%%%%%%%%%%%%%%%%
\begin{details}
\begin{itemize}
  \item If $\tail(e)=\sigma_i$ for some $1\leq i \leq s$, then we define
  \begin{align*}
  \htheta_e\big(\vx_i\big) = \vx_i \cdot B^{-1} A_{i, e} = (\vm_i, \vk_i) \cdot B^{-1} A_{i, e};
  \end{align*}
  \item If $v = \tail(e)$ is an intermediate node in $\mV \setminus (S \cup \{\rho\})$, then define
  \begin{align*}
  \htheta_e\big(y_d,~d \in \ein(v)\big) = \theta_e\big(y_d,~d \in \ein(v)\big) = \sum_{d \in \ein(v)} a_{d, e} \cdot y_d,
  \end{align*}
\end{itemize}
where $y_d \in \Fq$ is the message transmitted on edge $d$ under the source message vector $\vm_S$ and the key vector $\vk_S$.
\end{details}
%%%%%%%%%%%%%%%%%%%%%%%%%%%%%%%
For each $e \in \mE$, we let $\vh_e$ be the global encoding vector for the secure network code $\hmbC$. Similar to \eqref{ge-vec-form}, we write
\begin{align}
\vh_e = \begin{bmatrix}
\vh_e^{\,(\sigma_1)} \\
\vh_e^{\,(\sigma_2)} \\
\vdots \\
\vh_e^{\,(\sigma_s)} \\
\end{bmatrix},
\end{align}
and by the mechanism of linear network coding (similar to \eqref{equ_global_en_vector-i}), we have
\begin{align}\label{equ_sec_global_en_vector-i}
\vh_{e}^{\,(\sigma_i)}
    =
    B^{-1} \cdot A_{\sigma_i}
    \cdot
    \big( I-A \big)^{-1}\cdot \vec{1}_e= B^{-1} \cdot \vg_{e}^{\,(\sigma_i)}, \quad \forall~i=1, 2, \cdots, s.
\end{align}
Then, we obtain that
\begin{align}\label{equ_sec_global_en_vector}
\vh_{e} = \begin{bmatrix}
\vh_e^{\,(\sigma_1)} \\
\vh_e^{\,(\sigma_2)} \\
\vdots \\
\vh_e^{\,(\sigma_s)} \\
\end{bmatrix}
=\begin{bmatrix}
    B^{-1} A_{\sigma_1} \\
    B^{-1} A_{\sigma_2} \\
    \vdots \\
    B^{-1} A_{\sigma_s}
  \end{bmatrix}
  \cdot
\big( I-A \big)^{-1}
\cdot
\vec{1}_e
=\hB^{-1}\cdot \begin{bmatrix}
    A_{\sigma_1} \\
    A_{\sigma_2} \\
    \vdots \\
    A_{\sigma_s}
  \end{bmatrix}
  \cdot
\big( I-A \big)^{-1}
\cdot
\vec{1}_e
= \hB^{-1} \cdot \vg_{e}
\end{align}
(cf.~\eqref{equ_hat_B_inverse} for $\hB^{-1}$). Thus, we have constructed an $(R-r, 1)$ linear network code $\hmbC = \big\{ \htheta_e: e \in \mE;~\hvarphi \big\}$ for the secure model $(\mN, f, r)$, where the decoding function $\hvarphi$ will be clear later. The computability and security conditions (i.e., the admissibility) of the code $\hmbC$ will be verified in the next subsection.

\subsection{Verification of the Computability and Security Conditions}\label{sec: LSNC-verification}

\noindent\textbf{Verification of the Computability Condition:}

At the sink node $\rho$, the messages $y_e$, $e \in \ein(\rho)$ are received. Further, we let
\begin{align*}
G_{\rho} = \begin{bmatrix}
\vg_e:~e \in \ein(\rho)
\end{bmatrix}
\quad  \text{and} \quad
H_{\rho} =
\begin{bmatrix}
\vh_e:~e \in \ein(\rho)
\end{bmatrix}.
\end{align*}
Clearly, we have $H_{\rho} = \hB^{-1} \cdot G_{\rho}$ by \eqref{equ_sec_global_en_vector}, and then
\begin{align}\label{equ_decoding-equ}
\begin{split}
\vy_{\rho}\triangleq \big( y_e:~e \in \ein(\rho)\big)
& = \big( \vx_S \cdot \vh_e:~e \in \ein(\rho) \big) = \vx_S \cdot \begin{bmatrix} \vh_e:~e \in \ein(\rho)\end{bmatrix}\\
& = \vx_S \cdot H_{\rho} = \vx_S \cdot \hB^{-1} \cdot G_{\rho}
= \big( \vx_1 B^{-1}~~\vx_2 B^{-1} ~~ \cdots ~~ \vx_s B^{-1} \big) \cdot G_{\rho}.
\end{split}
\end{align}
Note that in \eqref{equ_decoding-equ}, $\vx_S$ is unknown while $\vy_{\rho}$ and $H_{\rho}$ (or equivalently, $\hB^{-1} \cdot G_{\rho}$) are known. We aim to compute $\sum_{i=1}^{s}\vm_i$ by ``solving'' the equation \eqref{equ_decoding-equ}.

In light of \eqref{equ_decoding-equ}, $\vy_{\rho}$ can also be regarded as the received vector at $\rho$ when we consider $\vx_i B^{-1}$ as the source messages generated by the source node $\sigma_i$ for $i=1,2,\cdots,s$ when using the linear network code~$\mbC$ for the model $(\mN, f)$. By the admissibility of the code $\mbC$ (i.e., the computability condition is satisfied for $\mbC$), we can apply the decoding function $\varphi$ of $\mbC$ on $\vy_{\rho}$ to compute with zero error
\begin{align}\label{apply_varphi_on_ye}
\sum_{i=1}^{s}\vx_i \cdot B^{-1} = \left( \sum_{i=1}^{s}\vx_i \right) \cdot B^{-1}.
\end{align}
%In fact, we recall from Appendix~\ref{appendix} that a linear decoding can be used to compute \eqref{apply_varphi_on_ye} with zero error.
Then, the admissibility of the linear network code $\mbC$ implies that
%that all the column vectors of the matrix $\left[\begin{smallmatrix}I_R \\ I_R \\ \vdots \\ I_R \end{smallmatrix}\right]_{Rs \times R}$ can be linearly represented by the column vectors of $G_{\rho}$
there exists an $|\ein(\rho)| \times R$ matrix $D$ in $\Fq$ such that\footnote{Here, we can take $D$ to be $K_{\rho}^\top$ in Appendix~\ref{appendix} and thus obtain \eqref{equ10}, i.e., \eqref{equ9-append}.}
\begin{align}\label{equ10}
  G_{\rho} \cdot D = \begin{bmatrix}
                       I_R \\
                       I_R \\
                       \vdots \\
                       I_R
                     \end{bmatrix}_{Rs \times R},
\end{align}
where $I_R$ stands for the $R\times R$ identity matrix. Immediately, we have
\begin{align}\label{computability_verified_matrix}
\vy_{\rho} D = \big( \vx_1 B^{-1}~~\vx_2 B^{-1} ~~ \cdots ~~ \vx_s B^{-1} \big) \cdot G_{\rho} \cdot D
=\sum_{i=1}^{s}\vx_i \cdot B^{-1}=\left(\sum_{i=1}^{s}\vx_i \right) \cdot B^{-1}.
\end{align}

By multiplying both sides of \eqref{apply_varphi_on_ye} by $B$, we obtain $\sum_{i=1}^{s}\vx_i$, or equivalently, $\Big (\sum_{i=1}^{s}\vm_i \quad  \sum_{i=1}^{s}\vk_i \Big)$. Thus, the $R-r$ function values
$$\sum_{i=1}^{s}\vm_i = \Big( \sum_{i=1}^{s} m_{i, j},~j=1, 2, \cdots, R-r \Big)$$
are computed at $\rho$ with zero error. We thus have verified the computability condition of the code $\hmbC$, and at the same time specified the decoding function $\hvarphi$, which indeed is linear. %Furthermore, for the admissible linear secure network code $\hmbC$ thus obtained, it has been shown that we are able to use a linear decoding function to compute the algebraic sum.}

\medskip

\noindent\textbf{Verification of the Security Condition:}

We let $\vX_i = (\vM_i~\vK_i )$, $i=1, 2, \cdots, s$ and $\vX_S = (\vX_1~\vX_2~\cdots~\vX_s)$. For each $e \in \mE$, we let $\vY_e = \vX_S \cdot \vh_e$, which is the random message transmitted on the edge $e$. We consider an arbitrary wiretap set $W \in \mW_r$ and will prove that
\begin{align}\label{security_condition_in_proof_1}
  H\big(\vM_S | \vY_W\big) = H\big (\vM_S\big ).
\end{align}
Toward this end, it suffices to prove that
\begin{align}\label{security_condition_in_proof_2}
  {\rm Pr}(\vM_S = \vm_S | \vY_W = \vy_W) = {\rm Pr}(\vM_S = \vm_S)
\end{align}
for any vector of source messages $\vm_S = (\vm_1, \vm_2, \cdots, \vm_s) \in \Fq^{(R-r)s}$ and any $\vy_W \triangleq (y_e:~e \in W) \in \Fq^{|W|}$ such that ${\rm Pr}\big(\vY_W = \vy_W\big)>0$.
%satisfying that there exists a vector of source messages $\vm'_S = (\vm'_1, \vm'_2, \cdots, \vm'_s) \in \Fq^{(R-r)s}$ and a vector of keys $\vk'_S = (\vk'_1, \vk'_2, \cdots, \vk'_s) \in \Fq^{rs}$ such that
%\begin{align*}
%\vy_W=\vx'_S\cdot
%\begin{bmatrix}
%    \vh_{e}:~e\in W
%  \end{bmatrix},
%\quad \text{ or equivalently, } \quad
%  y_e = \vx'_S \cdot \vh_e,\quad  \forall~e \in W,
%\end{align*}
%where $\vx'_S \triangleq (\vx'_1~ \vx'_2~ \cdots~ \vx'_s)$ with $\vx'_i = (\vm'_i~~ \vk'_i)$ for $i=1, 2, \cdots, s$.\footnote{Note that for such a vector $\vy_W$, we have ${\rm Pr}\big(\vY_W = \vy_W\big)>0$.}

First, we readily see that
\begin{align}\label{equ1-prob}
  {\rm Pr}\big(\vM_S = \vm_S\big) = \frac{1}{q^{(R-r)s}}, \quad \forall~\vm_S \in \Fq^{(R-r)s}.
\end{align}
Next, we consider
\begin{align}
&{\rm Pr}\big( \vM_S = \vm_S | \vY_W=\vy_W)\nonumber\\
&=\frac{{\rm Pr}\big( \vM_S = \vm_S,~ \vY_W=\vy_W)}{{\rm Pr}\big( \vY_W=\vy_W)}\nonumber\\
&=\frac{\sum_{\widehat{\vk}_S \in \Fq^{rs}} {\rm Pr}\big( \vM_S=\vm_S,~\vK_S=\widehat{\vk}_S,~\vY_W=\vy_W)}
{\sum_{\widehat{\vm}_S \in \Fq^{(R-r)s},~\widehat{\vk}_S \in \Fq^{rs}}
{\rm Pr}\big(\vM_S=\widehat{\vm}_S,~\vK_S=\widehat{\vk}_S,~\vY_W=\vy_W)}\label{equ1-veri_sec_cond}\\
&=\frac{\sum_{\widehat{\vk}_S \in\Fq^{rs}}{\rm Pr}\big(\vY_W=\vy_W|\vM_S=\vm_S,~\vK_S=\widehat{\vk}_S)\cdot{\rm Pr}\big(\vM_S=\vm_S,~\vK_S=\widehat{\vk}_S)}
{\sum_{\widehat{\vm}_S \in \Fq^{(R-r)s},~\widehat{\vk}_S \in \Fq^{rs}}
{\rm Pr}\big(\vY_W=\vy_W|\vM_S=\widehat{\vm}_S,~\vK_S=\widehat{\vk}_S)\cdot{\rm Pr}\big(\vM_S=\widehat{\vm}_S,~\vK_S=\widehat{\vk}_S\big)}\nonumber\\
&=\frac{\sum\limits_{\widehat{\vk}_S \in \Fq^{rs} \text{ s.t. } (\vm_S \ \widehat{\vk}_S) \cdot H_W=\vy_W}1\cdot{\rm Pr}\big(\vM_S=\vm_S,~\vK_S=\widehat{\vk}_S \big)}
{\sum\limits_{\widehat{\vm}_S \in \Fq^{(R-r)s},~\widehat{\vk}_S \in \Fq^{rs} \text{ s.t. } \atop (\widehat{\vm}_S \ \widehat{\vk}_S) \cdot H_W=\vy_W} 1 \cdot {\rm Pr}\big(\vM_S=\widehat{\vm}_S,~\vK_S=\widehat{\vk}_S \big)}\label{equ2-veri_sec_cond}\\
&=\frac{\#\big\{ \widehat{\vk}_S \in \Fq^{rs}:~ (\vm_S \ \widehat{\vk}_S) \cdot H_W=\vy_W \big\}}
{\#\big\{(\widehat{\vm}_S~\widehat{\vk}_S)\in \Fq^{(R-r)s}\times \Fq^{rs}:~(\widehat{\vm}_S \ \widehat{\vk}_S) \cdot H_W=\vy_W \big\}},\label{equ3-veri_sec_cond}
\end{align}
where in~\eqref{equ1-veri_sec_cond}, $\widehat{\vm}_S = (\widehat{\vm}_1~\widehat{\vm}_2~\cdots~\widehat{\vm}_s) \in \Fq^{(R-r)s}$ is a row vector with $\widehat{\vm}_i$ being a row $(R-r)$-vector for $i=1, 2, \cdots, s$ and similarly, $\widehat{\vk}_S = (\widehat{\vk}_1~\widehat{\vk}_2~ \cdots~ \widehat{\vk}_s) \in \Fq^{rs}$ is a row vector with $\widehat{\vk}_i$ being a row $r$-vector for $i=1, 2, \cdots, s$; in~\eqref{equ2-veri_sec_cond}, we let
\begin{align*}
  H_W = \begin{bmatrix}
    \vh_{e}:~e\in W
  \end{bmatrix}
\end{align*}
and the equality \eqref{equ2-veri_sec_cond} follows from
\begin{align*}
{\rm Pr}\big(\vY_W=\vy_W|\vM_S=\widehat{\vm}_S,~\vK_S=\widehat{\vk}_S)=
\begin{cases}
1, & \text{if $(\widehat{\vm}_S \ \widehat{\vk}_S) \cdot H_W=\vy_W$,}\\
0, & \text{otherwise;}
\end{cases}
\end{align*}
and in \eqref{equ3-veri_sec_cond}, we use ``$\#\{\cdot\}$'' to denote the cardinality of the set.

Further, we let $\Rank(H_W) = r'$. Clearly, $r' \leq |W| \leq r$. For the denominator of \eqref{equ3-veri_sec_cond}, it is easy to calculate that
\begin{align}\label{equ4-veri_sec_cond}
\#\big\{(\widehat{\vm}_S~\widehat{\vk}_S)\in \Fq^{(R-r)s}\times \Fq^{rs}:~(\widehat{\vm}_S \ \widehat{\vk}_S) \cdot H_W=\vy_W \big\} = q^{Rs-r'}.
\end{align}
Next, we focus on the numerator of~\eqref{equ3-veri_sec_cond}, where we note that $(\vm_S \ \widehat{\vk}_S)=\big((\vm_1~\widehat{\vk}_1)~(\vm_2~\widehat{\vk}_2)~ \cdots~(\vm_s~\widehat{\vk}_s)\big)$.
We denote by $I_R^{(R-r)}$ the submatrix of the $R \times R$ identity matrix $I_R$ consisting of its first $R-r$ columns. We further define the $Rs \times (R-r)s$ matrix
\begin{align*}
 \Gamma = \big[ I_R^{(R-r)}~~I_R^{(R-r)}~~\cdots~~I_R^{(R-r)} \big]_s^{\text{diagonal}}=
  \begin{bmatrix}
   I_R^{(R-r)} & \mathbf{0} & \cdots & \mathbf{0} \\
    \mathbf{0} & I_R^{(R-r)} & \cdots & \mathbf{0} \\
    \cdots & \cdots & \cdots & \cdots \\
    \mathbf{0} & \mathbf{0} & \cdots & I_R^{(R-r)}
  \end{bmatrix}_{Rs \times (R-r)s},
\end{align*}
where $\bfzero$ stands for the $R \times (R-r)$ zero matrix. Further, we can see that
\begin{align}
  & \#\big\{ \widehat{\vk}_S \in \Fq^{rs}:~ (\vm_S \ \widehat{\vk}_S) \cdot H_W=\vy_W \big\}\nonumber\\
  & = \#\big\{ (\widehat{\vm}_S~\widehat{\vk}_S) \in \Fq^{(R-r)s} \times \Fq^{rs}:~
  (\widehat{\vm}_S~ \widehat{\vk}_S) \cdot \Big[H_W~~\Gamma \Big] = (\vy_W~\vm_S) \big\}.\label{solution_number}
\end{align}
It thus suffices to calculate the number of solutions $(\widehat{\vm}_S~\widehat{\vk}_S)$ of the equation
\begin{align*}
   (\widehat{\vm}_S~ \widehat{\vk}_S) \cdot \Big[H_W~~\Gamma \Big] = (\vy_W~ \vm_S).
\end{align*}
Toward this end, we write
\begin{align}\label{rank_equal}
\Big[H_W~~\Gamma \Big] = \Big[ \hB^{-1} \cdot G_W~~\hB^{-1}\cdot\hB\cdot\Gamma \Big]=\hB^{-1} \cdot \Big[ G_W~~\hB\cdot\Gamma \Big],
\end{align}
where $G_W = \begin{bmatrix}\vg_{e}:~e\in W\end{bmatrix}$. We let $B^{(R-r)}$ be the $R \times (R-r)$ submatrix of $B$ consisting of its first $R-r$ columns, i.e.,
\begin{align*}
  B^{(R-r)} =\left[ \vb_1 ~~ \vb_2 ~~ \cdots ~~ \vb_{R-r} \right].
\end{align*}
Then, we see that
\begin{align*}
  \hB \cdot \Gamma = \Big[ B^{(R-r)}~~B^{(R-r)}~~\cdots~~B^{(R-r)} \Big]_s^{\text{diagonal}}
\end{align*}
and so
\begin{align}
  \Big[G_W~~\hB\cdot \Gamma \Big] & =\bigg[~\begin{bmatrix}\vg_{e}:~e\in W\end{bmatrix}~~  \Big[ B^{(R-r)}~~B^{(R-r)}~~\cdots~~B^{(R-r)} \Big]_s^{\text{diagonal}}~\bigg] \nonumber \\
  & =
  \begin{bmatrix}
    \vg_e^{\,(\sigma_1)}:~e \in W & B^{(R-r)} & \mathbf{0} & \cdots & \mathbf{0}\\
    \vg_e^{\,(\sigma_2)}:~e \in W & \mathbf{0} &  B^{(R-r)} & \cdots & \mathbf{0} \\
    \cdots & \cdots & \cdots & \cdots & \cdots \\
    \vg_e^{\,(\sigma_s)}:~e \in W & \mathbf{0} & \mathbf{0} &  \cdots & B^{(R-r)}
  \end{bmatrix}.\label{big_matrix_GW_BT}
\end{align}
Further, we recall that the $R-r$ vectors $\vb_1, \vb_2, \cdots,\vb_{R-r}$ are linearly independent and for each wiretap set $W \in \mW_r$, we have
\begin{align*}
\big\langle  \vb_j:~ 1\leq j \leq R-r  \big\rangle \bigcap
\big\langle \vg_e^{\,(\sigma_i)}:~ e \in W  \big\rangle=\{\bzero\},
\qquad  \forall~i = 1, 2, \cdots, s
\end{align*}
(cf.~\eqref{equ-sec-condition}).
Together with \eqref{rank_equal}, \eqref{big_matrix_GW_BT} and the invertibility of $\hB$, this implies that
\begin{align*}
 \Rank \Big[H_W~~\Gamma\Big] &= \Rank \Big[G_W~~\hB\cdot \Gamma\Big]= \Rank\big(G_W\big) + \Rank\big(\hB\cdot\Gamma\big) \\
                             &= \Rank\big(H_W\big) + s \cdot \Rank\big(B^{(R-r)}\big)
 =r' + (R-r)s.
\end{align*}
We thus obtain that
\begin{align*}
\#\Big\{ (\widehat{\vm}_S~\widehat{\vk}_S) \in \Fq^{(R-r)s} \times \Fq^{rs}:~
  (\widehat{\vm}_S~ \widehat{\vk}_S) \cdot \Big[H_W~~\Gamma \Big] = (\vy_W~\vm_S) \Big\}=q^{Rs- \Rank \big[H_W~~\Gamma\big]}= q^{rs-r'},
\end{align*}
and also by~\eqref{solution_number},
\begin{align}\label{number_hk_S}
 \#\big\{ \widehat{\vk}_S \in \Fq^{rs}:~ (\vm_S \ \widehat{\vk}_S) \cdot H_W=\vy_W \big\} =  q^{rs-r'}.
\end{align}
Combining \eqref{equ4-veri_sec_cond} and \eqref{number_hk_S} with \eqref{equ3-veri_sec_cond}, we immediately obtain that
\begin{align}\label{security_probability}
  {\rm Pr}\big(\vM_S = \vm_S | \vY_W = \vy_W\big) = \frac{1}{q^{(R-r)s}}.
\end{align}
Comparing \eqref{equ1-prob} and \eqref{security_probability}, we have proved
the equality~\eqref{security_condition_in_proof_2} and thus verified the security condition.

\subsection{The Required Field Size of the Code Construction}\label{sec: LSNC-field_size}

In this subsection, we present upper bounds on the minimum required field size of the existence of an admissible $(R-r, 1)$ linear secure network code $\hmbC$ for the secure model $(\mN, f, r)$ with $0\leq r \leq R \leq C_{\min}$. Following the upper bounds, we further show that by our code construction, for the algebraic sum $f$ over any finite field $\Fq$, we can always construct an $\Fq$-valued admissible (vector-) linear secure network code of rate up to $C_{\min}-r$ for the model $(\mN, f, r)$ with security level $0\leq r\leq C_{\min}$.

By applying the notion of primary minimum cut, we first prove in the next theorem a non-trivial upper bound on the minimum required field size of the existence of an admissible linear secure network code. This upper bound, which is graph-theoretic, only depends on the network topology and the required security level. Before presenting the upper bound, we recall that for an edge subset $W$, $\hW$ denotes the primary minimum cut separating $W$ from $D_W$ and $W=\hW$ is the condition for $W$ to be primary.

\begin{thm}\label{thm-enhanced-field-size}
Consider the model of secure network function computation $(\mN, f, r)$, where the target function $f$ is the algebraic sum over a finite field $\Fq$ and the security level $r$ satisfies $0 \leq r \leq C_{\min}$. Let
 \begin{align*}
\mW^*_r = \big\{W \in \mW_r:~ W=\hW \text{ with } |W|=r \big\}.
\end{align*}
Then, for any nonnegative integer $R$ with $r \leq R \leq C_{\min}$, there exists an $\Fq$-valued admissible $(R-r, 1)$ linear secure network code for $(\mN, f, r)$ if the field size $q$ satisfies
\begin{align}\label{enhanced-upperbound-fieldsize}
q> s\cdot|\mW^*_r|.
\end{align}
\end{thm}
\begin{IEEEproof}
Consider an arbitrary admissible secure network code $\hmbC$ for the model $(\mN, f, \mW^*_r)$. We first claim that the code $\hmbC$ is also admissible for the model $(\mN, f, r)$. It is evident that $\hmbC$ satisfies the computability condition for $(\mN, f, r)$. Then, it suffices to prove that $\hmbC$ satisfies the security condition for $(\mN, f, r)$, i.e., $I(\vY_W ; \vM_S)=0$, $\forall~W\in \mW_r$ provided that $I(\vY_W ; \vM_S)=0$, $\forall~W\in \mW^*_r$. Toward this end, we consider an arbitrary wiretap set $W\in \mW_r$ and let $W'$ be a cut separating $W$ from $D_W$ with $|W'|=\mincut(D_{W'},W')=r$, where $\mincut(D_{W'},W')$ denotes the minimum cut capacity separating $W'$ from $D_{W'}$. Further, we consider the primary minimum cut $\widehat{W'}$ separating~$W'$ from $D_{W'}$. By Lemma~\ref{prop1}, $\widehat{W'}$ is primary, or equivalently, $\widehat{W'}=\widehat{\widehat{W'}}$. Together with
$$\big|\widehat{W'}\big|=\mincut(D_{W'},W')=\big|W'\big|=r,$$
we have $\widehat{W'} \in \mW^*_r$ and thus $I(\vY_{\widehat{W'}} ; \vM_S)=0$ by the security condition of the code $\hmbC$ for $(\mN, f, \mW^*_r)$. This further implies $I(\vY_W ; \vM_S)=0$, because $\vY_W$ is a function of $\vY_{W'}$ and $\vY_{W'}$ is a function of $\vY_{\widehat{W'}}$ by the mechanism of network coding.

Next, we will consider the field size $q$ for constructing such an $\Fq$-valued admissible $(R-r, 1)$ linear secure network code for $(\mN, f, \mW^*_r)$ by applying the code construction in Section~\ref{sec: LSNC-construction}. First, let $\mbC$ be an $\Fq$-valued admissible $(R, 1)$ linear network code for the model $(\mN,f)$ of which all the global encoding vectors are
\begin{align*}
\vg_e=\left[\begin{smallmatrix}\vg_e^{\,(\sigma_1)} \\ \vg_e^{\,(\sigma_2)} \\ \vdots \\ \vg_e^{\,(\sigma_s)}\end{smallmatrix}\right],\quad e\in \mE.
\end{align*}
Similar to the verification of the security condition in Section~\ref{sec: LSNC-verification}, we readily see that an $\Fq$-valued admissible $(R-r, 1)$ linear secure network code for the secure model $(\mN, f, \mW^*_r)$ can be constructed if there exist $R$ linearly independent $\Fq$-valued column $R$-vectors $\vb_1, \vb_2, \cdots, \vb_{R}$ such that the first $R-r$ vectors $\vb_1, \vb_2, \cdots, \vb_{R-r}$ satisfy the condition that for all the wiretap sets $W \in \mW^*_r$,
\begin{align}\label{thm-field-size_equ-sec-condition}
\big\langle  \vb_j:~ 1\leq j \leq R-r  \big\rangle \bigcap
\big\langle \vg_e^{\,(\sigma_i)}:~ e \in W  \big\rangle=\{\bzero\}, \quad \forall~i = 1, 2, \cdots, s.
\end{align}
In the following, we will prove that such $R$ vectors $\vb_1, \vb_2, \cdots, \vb_{R}$ exist if the inequality~\eqref{enhanced-upperbound-fieldsize} holds. The techniques involved here are standard in the literature.

For notational simplicity, we let
\begin{align*}
\mB_j=\big\langle \vb_1,~\vb_2,~\cdots,~\vb_{j-1} \big\rangle, \quad \forall~ 1 \leq j \leq R,
\end{align*}
and
\begin{align*}
\mL_W^{(\sigma_i)}=\big\langle  \vg_e^{\,(\sigma_i)}:~e\in W \big\rangle, \quad \forall~ 1 \leq i \leq s ~\text{ and }~ W\in \mW^*_r.
\end{align*}
Now, we choose the $R$ vectors $\vb_1$, $\vb_2$, $\cdots$, $\vb_R$ in $\Fq^R$ sequentially as follows:
\begin{itemize}
  \item For $1 \leq j \leq R-r$, we choose
\begin{align}\label{vb_j-1}
\vb_j \in
\Fq^R \setminus \bigcup_{W \in \mW^*_r}~\bigcup_{i=1}^s \big(\mL_W^{(\sigma_i)} + \mB_{j-1}  \big);
\end{align}
  \item For $R-r+1 \leq j \leq R$, we choose
\begin{align}\label{vb_j-2}
\vb_j \in
\Fq^R \setminus \mB_{j-1}.
\end{align}
\end{itemize}
By~\eqref{vb_j-1} and~\eqref{vb_j-2}, we have
\begin{align*}
\vb_1 \neq \bzero \quad \text{ and }\quad  \vb_j \in \Fq^R \setminus \mB_{j-1},~~2\leq j \leq R,
\end{align*}
which implies the linear independence of $\vb_1, \vb_2, \cdots, \vb_{R}$. On the other hand, it follows from~\eqref{vb_j-1} that the condition~\eqref{thm-field-size_equ-sec-condition} is satisfied for each wiretap set $W\in \mW^*_r$.

It now remains to prove that all the sets on the right hand sides of~\eqref{vb_j-1} and \eqref{vb_j-2} are nonempty provided that $q>s\cdot|\mW^*_r|$. For $1\leq j \leq R-r$, by~\eqref{vb_j-1}, we obtain that
\begin{align}
&\bigg|\Fq^R \setminus \bigcup_{W \in \mW^*_r}~\bigcup_{i=1}^s \big(\mL_W^{(\sigma_i)} + \mB_{j-1}  \big) \bigg|\nonumber\\
&=\bigg|\Fq^R \bigg|-\bigg| \bigcup_{W \in \mW^*_r}~\bigcup_{i=1}^s \big(\mL_W^{(\sigma_i)} + \mB_{j-1}  \big) \bigg|\nonumber\\
&\geq q^R-\sum_{W \in \mW^*_r} \sum_{i=1}^s  \Big| \mL_W^{(\sigma_i)} + \mB_{j-1}  \Big|\nonumber\\
& \geq q^R-\sum_{W \in \mW^*_r} \sum_{i=1}^s q^{R-1}\label{equ1-pf-thm-field-size}\\
& =q^{R-1} \big( q-s\cdot|\mW^*_r| \big)>0,\nonumber
\end{align}
where the inequality \eqref{equ1-pf-thm-field-size} holds because for any $1\leq i \leq s$, $1\leq j \leq R-r$ and $W\in\mW^*_r$, we have
\begin{align*}
\dim\big( \mL_W^{(\sigma_i)} \big) \leq r \quad \text{ and } \quad
\dim\big( \mB_{j-1} \big) \leq R-r-1,
\end{align*}
implying that
\begin{align*}
\dim\big( \mL_W^{(\sigma_i)} + \mB_{j-1} \big)\leq \dim\big( \mL_W^{(\sigma_i)} \big)+
\dim\big( \mB_{j-1} \big) \leq r+R-r-1=R-1.
\end{align*}
For $R-r+1 \leq j \leq R$, by~\eqref{vb_j-2} we have
\begin{align*}
\Big|\Fq^R \setminus \mB_{j-1} \Big|=q^R-q^{j-1} \geq q^R-q^{R-1}>0.
\end{align*}
The theorem is thus proved.
\end{IEEEproof}

\medskip

%\begin{thm}\label{thm-field-size}
%Consider the model of secure network function computation $(\mN, f, r)$, where the target function $f$ is the algebraic sum over a finite field $\Fq$ and the security level $r$ satisfies $0 \leq r \leq C_{\min}$. Then, for any nonnegative integer $R$ with $r \leq R \leq C_{\min}$, there exists an $\Fq$-valued admissible $(R-r, 1)$ linear secure network code for $(\mN, f, r)$ if the field size $q$ satisfies
%\begin{align}\label{upperbound-fieldsize}
%q> s\cdot|\mW_r|.
%\end{align}
%\end{thm}

We can readily see that
\begin{align*}
\big|\mW^*_r\big| \leq \Big| \big\{ W \subseteq \mE:~\vert W \vert = r \big\}\Big| = {|\mE| \choose r}.
\end{align*}
Following Theorem~\ref{thm-enhanced-field-size}, this immediately gives an upper bound on the required field size in closed form that does not depend on the network topology.

\begin{cor}\label{cor-upper-bound-field-size}
Consider the model of secure network function computation $(\mN, f, r)$, where the target function $f$ is the algebraic sum over a finite field $\Fq$ and the security level $r$ satisfies $0 \leq r \leq C_{\min}$.
Then, for any nonnegative integer $R$ with $r \leq R \leq C_{\min}$, there exists an $\Fq$-valued admissible $(R-r, 1)$ linear secure network code for $(\mN, f, r)$ if the field size $q$ satisfies $q> s\cdot {|\mE| \choose r}$.
\end{cor}

\begin{remark}
For the case that the security level $r=0$, i.e., to compute the algebraic sum $f$ over the network $\mN$ without security constraint, we have $|\mW^*_0|=1$ since $\mW_0=\mW^*_0=\{\emptyset\}$. On the other hand, ${|\mE| \choose 0}=1$. Thus, the upper bounds in Theorem~\ref{thm-enhanced-field-size} and Corollary~\ref{cor-upper-bound-field-size} are both equal to $s$, the number of source nodes.
%This guarantees the existence of an admissible $(R,1)$ linear network code for the model $(\mN, f)$ for any rate $0\leq R \leq C_{\min}$. To be specific, in our code construction, for any $0\leq R \leq C_{\min}$, an admissible $(R,1)$ linear network code for $(\mN, f)$ can be constructed by ``reversing'' a rate-$R$ linear network code for the network coding problem on the reversed graph $\mG^\top$ such that the single source node $\rho$ multicasts messages to the $s$ sink nodes $\sigma_1, \sigma_2,\cdots, \sigma_s$ in $S$. The field size $q > s$ guarantees the existence of such a linear network code for the single-source multicast problem (cf.~\cite{Yeung-book}).

%For the case of the security level $1\leq r \leq R \leq C_{\min}$, both the upper bounds are strictly larger than $s$. This similarly guarantees that we are able to construct an admissible $(R,1)$ linear network code for $(\mN, f)$, as a building block to construct an admissible $(R-r, 1)$ linear secure network code for $(\mN, f, r)$.
\end{remark}

We continue to consider the secure model $(\mN, f, r)$, where the target function $f$ is the algebraic sum on a finite field $\Fq$ and the security level $r\leq C_{\min}$. For the case that the field size $q > s\cdot|\mW_r^*|$, by Theorem~\ref{thm-enhanced-field-size}, we can construct an $\Fq$-valued admissible $(R-r, 1)$ linear secure network code for any rate $1 \leq R\leq C_{\min}$ by using our code construction. Next, we consider the case that the field size $q \leq s\cdot|\mW_r^*|$. We take an extension field $\FqL$ of $\Fq$ such that the size of the extension field $q^L>s\cdot|\mW_r^*|$. Then, an $\FqL$-valued admissible $(R-r, 1)$ linear secure network code can be obtained by our code construction. Note that the extension field $\FqL$ can be viewed as an $L$-dimensional vector space over $\Fq$ with the basis $\{1, \alpha, \alpha^2, \cdots, \alpha^{L-1} \}$, where $\alpha$ is a primitive element of $\FqL$. Hence, the $\FqL$-valued $(R-r, 1)$ linear secure network code for computing the algebraic sum on $\FqL$ can be regarded as an $\big( (R-r)L, L \big)$ linear secure network code for computing the algebraic sum $f$ on $\Fq$, which also has the same secure computing rate $R-r$. Combining the two cases, for the algebraic sum $f$ over any finite field $\Fq$, we can always construct an $\Fq$-valued admissible (vector-) linear secure network code of rate up to $C_{\min}-r$ for the model $(\mN, f, r)$ with security level $0\leq r\leq C_{\min}$.

\subsection{An Example}

In this subsection, we will give an example to illustrate our code construction. Furthermore, for the model considered in this example, there exists another admissible linear secure network code of the same (optimal) rate on a smaller field that cannot be obtained by our code construction.
%Thus, another systematic approaches of code construction are deserved to investigate.

\begin{example}\label{eg}

\begin{figure}[t]
  \centering
{
 \begin{tikzpicture}[x=0.6cm]
    \draw (-3,0) node[vertex] (1) [label=above:$\sigma_1$] {};
    \draw ( 3,0) node[vertex] (2) [label=above:$\sigma_2$] {};
    \draw ( 0,-1.5) node[vertex] (3) [label=left:] {};
    \draw (-3,-5) node[vertex] (4) [label=left:] {};
    \draw ( 3,-5) node[vertex] (5) [label=right:] {};
    \draw ( 0,-3.5) node[vertex] (6) [label=right:] {};
    \draw ( 0,-6.5) node[vertex] (7) [label=below: $\rho$] {};

    \draw[->,>=latex] (1) -- (4) node[midway, auto,swap, left=-1mm] {$e_1$};
    \draw[->,>=latex] (1) -- (3) node[midway, auto, right=-0.5mm] {$e_2$};
    \draw[->,>=latex] (2) -- (3) node[midway, auto,swap, left=-0.5mm] {$e_3$};
    \draw[->,>=latex] (2) -- (5) node[midway, auto, right=-1mm] {$e_4$};
    \draw[->,>=latex] (3) -- (6) node[midway, auto, right=-1mm] {$e_5$};
    \draw[->,>=latex] (6) -- (4) node[midway, auto, left=-0.5mm] {$e_6$};
    \draw[->,>=latex] (6) -- (5) node[midway, auto,swap, right=-0.5mm] {$e_7$};
    \draw[->,>=latex] (4) -- (7) node[midway, auto,swap, right=-0.5mm] {$e_8$};
    \draw[->,>=latex] (5) -- (7) node[midway, auto, left=-0.5mm] {$e_9$};
    \end{tikzpicture}
}
\caption{The butterfly network $\mN$.}
  \label{fig:butterfly_network}
\end{figure}

We consider a secure model $(\mN, f, r)$, where $\mN=(\mG,S,\rho)$ is the network with the butterfly graph $\mG$ depicted in Fig.~\ref{fig:butterfly_network}, the set of source nodes $S=\{\sigma_1,\sigma_2\}$, the target function $f$ is the algebraic sum over the finite field $\mathbb{F}_2$, and the security level $r=1$. For the network $\mN$, we readily see that $C_{\min}=\overline{C}_{\min}=2$ from
\begin{align*}
\{e_1,e_2\}\in \arg\min_{C}\big\{|C|:~C\in \Lambda(\mN) \text{ and } D_C=I_C \big\}.
\end{align*}
By Corollary~\ref{cor10}, we have
\begin{align*}
\hmC(\mN,f,r)=C_{\min}-r=1.
\end{align*}

In the following, we will construct an optimal linear secure network code $\hmbC$ (i.e., $R(\hmbC)=\hmC(\mN,f,r)=1$) for the secure model $(\mN, f, r)$ by our code construction. First, let $\mathbb{F}_4=\{ 0,1, \alpha, 1+\alpha\}$ be the extension field of degree $2$ over $\mathbb{F}_{2}$, where $\alpha$ is a primitive element of $\mathbb{F}_4$. Consider an (optimal) $\mathbb{F}_4$-valued $(2,1)$ linear network code $\mbC$ on the network $\mN$ for computing the algebraic sum over $\mathbb{F}_4$, of which all the global encoding vectors are
\begin{align}\label{global_kernels_EX}
\begin{split}
&\vg_{e_1}=\left[\begin{smallmatrix} 1 \\ 1 \\ 0 \\ 0 \end{smallmatrix}\right], \quad \vg_{e_2}=\left[\begin{smallmatrix}  0 \\ 1 \\ 0 \\ 0 \end{smallmatrix}\right], \quad
\vg_{e_3}=\left[\begin{smallmatrix}  0 \\ 0 \\ 1 \\ 0 \end{smallmatrix}\right], \quad
\vg_{e_4}=\left[\begin{smallmatrix}  0 \\ 0 \\ 1 \\ 1 \end{smallmatrix}\right], \\
&
\vg_{e_5}=\vg_{e_6}=\vg_{e_7}=\left[\begin{smallmatrix}  0 \\ 1 \\ 1 \\ 0 \end{smallmatrix}\right], \quad
\vg_{e_8}=\left[\begin{smallmatrix}  1 \\ 0 \\ 1 \\ 0 \end{smallmatrix}\right], \quad
\vg_{e_9}=\left[\begin{smallmatrix}  0 \\ 1 \\ 0 \\ 1 \end{smallmatrix}\right].
\end{split}
\end{align}
This code $\mbC$ can be regarded as an $\mathbb{F}_2$-valued $(4,2)$ linear network code on the network $\mN$ for computing the algebraic sum over $\mathbb{F}_2$, which is elaborated as follows. To be specific, we regard the extension field~$\mathbb{F}_4$ as a $2$-dimensional vector space over $\mathbb{F}_2$ with the basis $\{1, \alpha\}$. For $i=1,2$, the source node $\sigma_i$ sequentially generates $4$ symbols in $\mathbb{F}_2$, say, $m_{i,1}$, $m_{i,2}$, $m_{i,3}$, $m_{i,4}$, which are regarded as 2 elements in $\mathbb{F}_4$:
\begin{align*}
x_{i,1}\triangleq m_{i,1}+m_{i,2}\cdot \alpha \quad \text{ and } \quad x_{i,2}\triangleq m_{i,3}+m_{i,4}\cdot \alpha.
\end{align*}
Using the code $\mbC$, the sink node $\rho$ receives
\begin{align*}
y_{e_8}=( x_{1,1}~~x_{1,2}~~x_{2,1}~~x_{2,2} )\cdot \vg_{e_8}=x_{1,1}+x_{2,1}=(m_{1,1}+m_{2,1})+(m_{1,2}+m_{2,2})\cdot \alpha,\\
y_{e_9}=( x_{1,1}~~x_{1,2}~~x_{2,1}~~x_{2,2} )\cdot \vg_{e_9}=x_{1,2}+x_{2,2}=(m_{1,3}+m_{2,3})+(m_{1,4}+m_{2,4})\cdot \alpha.
\end{align*}
Thus, the 4 function values $f(m_{1,j},m_{2,j})=m_{1,j}+m_{2,j}$, $j=1,2,3,4$ are computed with zero error at the sink node $\rho$.

Based on the code $\mbC$, we will construct an $\mathbb{F}_4$-valued $(1,1)$ linear secure network code $\hmbC$ on the network $\mN$ for securely computing the algebraic sum over $\mathbb{F}_4$ with the security level $r=1$. This code $\hmbC$ can also be similarly regarded as an $\mathbb{F}_2$-valued $(2,2)$ linear secure network code for $(\mN,f,r)$. By our construction, it suffices to construct an $\mathbb{F}_4$-valued $2 \times 2$ invertible matrix $B = \left[ \vb_1 ~~ \vb_2 \right]$ such that for any edge $e\in \mE$, or equivalently, for any wiretap set $\{e\} \in  \mW_1$,
\begin{align*}
\big\langle  \vb_1  \big\rangle \bigcap
\big\langle \vg_e^{\,(\sigma_i)} \big\rangle=\{\bzero\},
\qquad  \forall~i = 1, 2,
\end{align*}
(cf.~\eqref{equ-sec-condition}). To satisfy the above condition, we choose
\begin{align*}
\vb_1=\begin{bmatrix} 1 \\ \alpha \end{bmatrix} \quad \text{ and }\quad  \vb_2=\begin{bmatrix} 0 \\ 1 \end{bmatrix},
\end{align*}
i.e.,
\begin{align*}
 B=\Big[ \vb_1 \ \  \vb_2 \Big]=\begin{bmatrix} 1 & 0  \\ \alpha & 1 \end{bmatrix}.
\end{align*}
Then,
$B^{-1}=B=\Big[\begin{smallmatrix} 1 & 0  \\ \alpha & 1 \end{smallmatrix}\Big]$.
We further let
\begin{align*}
\hB = \big[ B~~B  \big]_2^{\text{\rm diagonal}}
=
\begin{bmatrix}
B & \bfzero \\
\bfzero & B \\
\end{bmatrix},
\end{align*}
and then $\hB^{-1}\triangleq \big[ B^{-1}~~B^{-1}  \big]_2^{\text{\rm diagonal}}=\hB$. We thus have obtained
an $\mathbb{F}_4$-valued admissible $(1,1)$ linear secure network code $\hmbC$ on the network $\mN$ for securely computing the algebraic sum over $\mathbb{F}_4$ with security level $r=1$. The global encoding vectors of $\hmbC$ are $\vh_{e}=\hB^{-1}\cdot \vg_e$, $e\in \mE$, i.e.,
\begin{align}\label{sec_global_kernels_EX}
\begin{split}
&\vh_{e_1}=\left[\begin{smallmatrix} 1 \\ 1+\alpha \\ 0 \\ 0 \end{smallmatrix}\right], \quad \vh_{e_2}=\left[\begin{smallmatrix}  0 \\ 1 \\ 0 \\ 0 \end{smallmatrix}\right], \quad
\vh_{e_3}=\left[\begin{smallmatrix}  0 \\ 0 \\ 1 \\ \alpha \end{smallmatrix}\right], \quad
\vh_{e_4}=\left[\begin{smallmatrix}  0 \\ 0 \\ 1 \\ 1+\alpha \end{smallmatrix}\right], \\
&
\vh_{e_5}=\vh_{e_6}=\vh_{e_7}=\left[\begin{smallmatrix}  0 \\ 1 \\ 1 \\ \alpha \end{smallmatrix}\right], \quad
\vh_{e_8}=\left[\begin{smallmatrix}  1 \\ \alpha \\ 1 \\ \alpha \end{smallmatrix}\right], \quad
\vh_{e_9}=\left[\begin{smallmatrix}  0 \\ 1 \\ 0 \\ 1 \end{smallmatrix}\right].
\end{split}
\end{align}

In using the obtained code $\hmbC$, for $i=1,2$, let $m_{i,1}$ and $m_{i,2}$ in $\mathbb{F}_2$ be two symbols generated by the source node $\sigma_i$, which can be regarded as the element $\vm_i\triangleq m_{i,1}+m_{i,2}\cdot \alpha $ in $\mathbb{F}_4$. Further, let $\vk_1\in \mathbb{F}_4$ and $\vk_2\in \mathbb{F}_4$ be two arbitrary outputs of the random keys $\vK_1$ and $\vK_2$, respectively, where $\vK_1$ and $\vK_2$ are two i.i.d. random variables with the uniform distribution on $\mathbb{F}_4$. Then, we let
$\vx_1=(\vm_1~\vk_1)$, $\vx_2=(\vm_2~\vk_2)$, and $\vx_S=(\vx_1~\vx_2)=(\vm_1~\vk_1~\vm_2~\vk_2)$.

We use $y_e$, which takes values in $\mathbb{F}_4$, to denote the message transmitted on each edge $e\in\mE$. By the global encoding vectors of $\hmbC$ ({\rm cf}.~\eqref{sec_global_kernels_EX}), the messages $y_e=\vx_S \cdot \vh_e$ transmitted on the edges $e\in\mE$ are
\begin{align}\label{equ-algebraic_sum_keys}
\begin{split}
&y_{e_1}=\vm_1+ (1+\alpha) \cdot \vk_1, \quad y_{e_2}=\vk_1,\quad y_{e_3}=\vm_2+ \alpha \cdot \vk_2, \quad y_{e_4}=\vm_2+ (1+\alpha) \cdot \vk_2,\\
&y_{e_5}=y_{e_6}=y_{e_7}=\vk_1 + \vm_2 + \alpha \cdot \vk_2,\quad y_{e_8}=\vm_1+ \alpha\cdot\vk_1 + \vm_2+\alpha\cdot \vk_2,\quad y_{e_9}=\vk_1+\vk_2.
\end{split}
\end{align}
We can readily verify the computability and security conditions for the code $\hmbC$. More precisely, by the two messages $y_{e_8}$ and $y_{e_9}$ received at the sink node $\rho$, the algebraic sum $\vm_1+\vm_2$ over $\mathbb{F}_4$, or equivalently, the algebraic sums $m_{1,1}+m_{2,1}$ and $m_{1,2}+m_{2,2}$ over $\mathbb{F}_2$, are computed with zero error at the sink node~$\rho$. On the other hand, it is also easy to check that the wiretapper cannot obtain any information
about the source messages $\vm_1$ and $\vm_2$ when any one edge is eavesdropped.
\end{example}

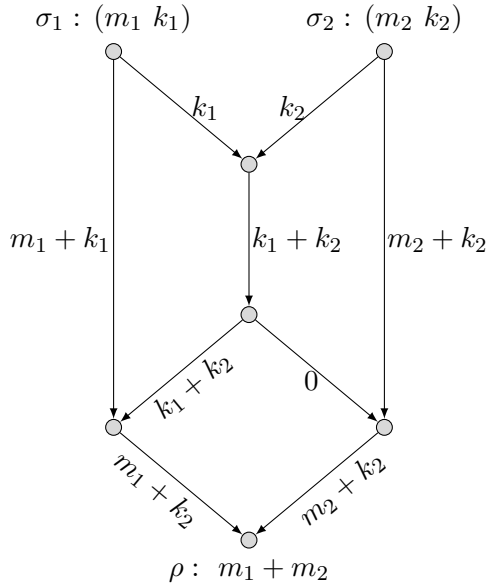
\begin{figure}[t]
\centering
 \begin{tikzpicture}[x=0.6cm]
    \draw (-3,0) node[vertex] (1) [label=above:{$\sigma_1:\,(m_{1}~k_{1})$}] {};
    \draw ( 3,0) node[vertex] (2) [label=above:{$\sigma_2:\,(m_{2}~k_{2})$}] {};
    \draw ( 0,-1.5) node[vertex] (3) [label=left:] {};
    \draw (-3,-5) node[vertex] (4) [label=left:] {};
    \draw ( 3,-5) node[vertex] (5) [label=right:] {};
    \draw ( 0,-3.5) node[vertex] (6) [label=right:] {};
    \draw ( 0,-6.5) node[vertex] (7) [label=below: $\rho:~m_{1}+m_{2}$] {};

    \draw[->,>=latex] (1) -- (4) node[midway, auto,swap, left=-1mm] {$m_{1}+k_{1}$};
    \draw[->,>=latex] (1) -- (3) node[midway, auto, right=-0mm] {$k_{1}$};
    \draw[->,>=latex] (2) -- (3) node[midway, auto,swap, left=-0mm] {$k_{2}$};
    \draw[->,>=latex] (2) -- (5) node[midway, auto, right=-1mm] {$m_{2}+k_{2}$};
    \draw[->,>=latex] (3) -- (6) node[midway, auto, right=-1mm] {$k_{1}+k_{2}$};
    \draw[->,>=latex] (6) -- (4) node[midway, sloped, below=-0.5mm] {$k_{1}+k_{2}$};
    \draw[->,>=latex] (6) -- (5) node[pos=0.6, auto, left=-0mm] {$0$};
    \draw[->,>=latex] (4) -- (7) node[pos=0.4, sloped, below=-0.3mm] {$m_{1}+k_{2}$};
    \draw[->,>=latex] (5) -- (7) node[pos=0.4, sloped, swap, below=-0.3mm] {$m_{2}+k_{2}$};
    \end{tikzpicture}
\caption{An $\mathbb{F}_2$-valued $(1,1)$ linear secure network code for the secure model $(\mN, f, r)$.}
  \label{butfly_net_rate_3_over_2}
\end{figure}

For the secure model $(\mN, f, r)$ as discussed in the above example, by Theorem~\ref{thm-enhanced-field-size}, we can construct an $\Fq$-valued admissible $(1,1)$ linear secure network code for computing the algebraic sum over $\Fq$ on the network $\mN$ with security level $r=1$ if the field size $q > 2\!\cdot\!|\mW_1^*|=14$, where $\mW_1^*=\{e_1,e_2,e_3,e_4,e_5,e_8,e_9\}$. However, we see in the example that the finite field $\mathbb{F}_4$ is sufficient for our code construction. This implies that for our code construction, the bound~\eqref{enhanced-upperbound-fieldsize} in Theorem~\ref{thm-enhanced-field-size} on the field size is only sufficient but far from being necessary.

We continue to consider the secure model $(\mN, f, r)$ in the above example. In Fig.~\ref{butfly_net_rate_3_over_2}, we show an $\mathbb{F}_2$-valued admissible $(1,1)$ linear secure network code which is also an optimal code achieving the secure computing capacity $\hmC(\mN,f,r)=1$. Note that this code requires only $\mathbb{F}_2$, which is smaller than~$\mathbb{F}_4$ required in the example. However, this code is yet to be obtained by our code construction. More precisely, according to the code construction in Section~\ref{sec: LSNC-construction}, we construct an admissible linear secure network code for the secure model $(\mN, f, r)$ by using a common linear transformation at all the source nodes on an admissible linear network code for $(\mN, f)$. As such, the algebraic sum of all the keys $\sum_{i=1}^{s}\vk_i$ can be always computed with zero error at the sink node $\rho$ (cf.~\ref{sec: LSNC-verification}), e.g., $\vk_1+\vk_2$ in the example (cf.~\eqref{equ-algebraic_sum_keys}). However, we can readily see that in Fig.~\ref{butfly_net_rate_3_over_2}, the code is not the case, where the key $k_1$ of the source node~$\sigma_1$ is cancelled out before arriving to $\rho$.

\section{Conclusion}\label{sec:concl}

Secure network function computation has been put forward in this paper. As the first work on this problem, we have investigated the special case of securely computing a linear function over a finite field with a wiretapper who can eavesdrop any subset of edges up to a certain size but is not allowed to obtain any information about the source messages. We have proved an upper bound on the secure computing capacity, which is applicable to arbitrary network topologies and arbitrary security levels. Since this upper bound is not in closed form, we also have proposed an efficient graph-theoretic approach to compute it in a linear time of the number of edges in the network. Furthermore, we have developed a code construction for linear function-computing secure network codes. Based on this construction, we have obtained a lower bound on the secure computing capacity and provided two sufficient conditions in terms of the network topology on the tightness of the lower bound. By combining the obtained upper and lower bounds, we have fully characterized the secure computing capacities for some classes of secure models.

For the model considered in the current paper, several interesting problems still remain open, such as whether other code constructions can be designed for smaller field sizes or to achieve higher secure computing rates, what the minimum size of the random key or the minimum entropy of the random key for each source node is, etc.
In the general setup of the model of secure network function computation, we can also consider different criteria for security. In Part~\Rmnum{2} of this paper, we will investigate securely computing a linear function over a finite field, where the wiretapper is not allowed to obtain any information about this linear function.

%%%%%%%%%%%%%%%%%%%%%%%%%%%%%%%%%%%%%%%%%%%%%%%%%
\numberwithin{thm}{section}
\appendices

%%%%%%%%%%%%%%%%%%%%%%%%%%%%%%%%%%%%%%%%%%%%%%%%%

\section{Construction of Linear Network Codes for the Model $(\mN,f)$}\label{appendix}

%In this appendix, we elaborate the construction of an admissible linear network code for the model $(\mN, f)$ by ``reversing'' a linear network code for the single-source multicast problem on the {\em reverse graph} $\mG^\top$ (we refer the reader to~\cite{Koetter-CISS2004,Rai-Dey-TIT-2012} for details) or (we refer the reader to Appendix for any nonnegative integer $R$ with $R \leq C_{\min}$, an $(R, 1)$

We refer the reader to the definition of an admissible $(R,1)$  linear network code for the model $(\mN,f)$ in Section~\ref{sec: LSNC-construction}, where some notations therein are adopted in this appendix.

Consider the single-source multicast problem on the reversed network $\mN^\top \triangleq (\mG^\top,S,\rho)$ of $\mN$, where the node~$\rho$, regarded as the single source node, is required to multicast the source message on the reversed graph $\mG^\top$ to the nodes $\sigma_i \in S$, regarded as all the sink nodes. On $\mG^{\top}$, the tail node and head node of an edge $e \in \mE$ are denoted by $\tail^{\mG^{\top}}\!(e)$ and $\head^{\mG^{\top}}\!(e)$, respectively. Clearly, we have $\tail^{\mG^{\top}}\!(e)=\head(e)$ and $\head^{\mG^{\top}}\!(e)=\tail(e)$. %, where we recall that $\head(e)$ and $\tail(e)$ are the head node and tail node of $e$ on the original graph $\mG$, respectively.
For a node $v$ in $\mG^{\top}$, we let $\ein^{\mG^{\top}}\!(v)$ and $\eout^{\mG^{\top}}\!(v)$ be the set of input edges and the set of output edges on the reversed graph $\mG^{\top}$, respectively. Then, we have
\begin{align}\label{equ1-append}
\ein^{\mG^{\top}}\!(v)=\eout(v) \quad \text{ and } \quad \eout^{\mG^{\top}}\!(v)=\ein(v).
\end{align}
For a node~$\sigma_i\in S$, we see that $\mincut^{\mG^\top}\!(\rho,\sigma_i)=\mincut(\sigma_i,\rho)$, where $\mincut^{\mG^\top}\!(\rho,\sigma_i)$ denotes the minimum cut capacity separating $\sigma_i$ from $\rho$ on $\mG^\top$. Thus, $$\min_{1\leq i \leq s}\mincut^{\mG^\top}\!(\rho,\sigma_i)=\min_{1\leq i \leq s}\mincut(\sigma_i,\rho),$$
which are both denoted by $C_{\min}$.

We now consider an $\Fq$-valued rate-$R$ (scalar) linear network code for the single-source multicast problem on the reversed network $\mN^\top$, where $\Fq$ is a finite field of order $q$ and the rate $R$ is a nonnegative integer not larger than $C_{\min}$, the theoretical maximum rate at which $\rho$ can multicast the source message to all the sink nodes in~$S$ on $\mN^\top$. %It has been proved that linear network coding over a finite field is sufficient for achieving the maximum rate~$C_{\min}$ \cite{Li-Yeung-Cai-2003, Koetter-Medard-algebraic}.
For the rate $R$, we introduce $R$ {\em imaginary source edges} connecting to~$\rho$, denoted by $d_1',d_2',\cdots,d_R'$, respectively, and let $\ein^{\mG^\top}\!(\rho)=\big\{d_1',d_2',\cdots,d_R' \big\}$. For the source message  $\bx=\big(x_1~x_2~\cdots~x_R \big) \in \Fq^{R}$ of $R$ source symbols generated by~$\rho$, we assume without loss of generality that $x_i$ is transmitted on the $i$th imaginary channel $d'_i$, $1\leq i \leq R$.

\begin{defn}\label{def-LNC}
For the single-source multicast problem on the reversed network $\mN^\top \triangleq (\mG^\top,S,\rho)$, an $\Fq$-valued rate-$R$ linear network code $\mbC_{\mN^\top}$ consists of a local encoding kernel $K_v$ for each non-sink node $v$ in $\mV\setminus S$ and a decoding matrix $F_{\sigma_i}$ for each sink node $\sigma_i\in S$, as specified below:
\begin{itemize}
  \item For each non-sink node $v \in \mV\setminus S$, the local encoding kernel is an $\Fq$-valued $\big\lvert \ein^{\mG^\top}\!(v) \big\rvert \times \big\lvert \eout^{\mG^\top}\!(v) \big\rvert$ matrix
      $$K_v=\big[k_{d,e}\big]_{d\in \ein^{\mG^\top}\!(v),\, e\in \eout^{\mG^\top}\!(v)},$$
      where $k_{d,e}\in \Fq$ is called the local encoding coefficient for the adjacent edge pair $(d,e)$;
  \item For each sink node $\sigma_i\in S$, the decoding matrix is an $\Fq$-valued $R \times \big\lvert \ein^{\mG^\top}\!(\sigma_i) \big\rvert$ matrix
  $$F_{\sigma_i}=\left[f_e:~e\in \ein^{\mG^\top}\!(\sigma_i)\right]$$
  induced by the local encoding kernels $K_v$, $v \in \mV\setminus S$, where for each edge $e$ in $\mE$, $f_e$ is a column $R$-vector, called the global encoding kernel of~$e$, which can be calculated recursively according to the reverse order of ``$\prec$'' (cf.~the second paragraph of Section~\ref{sec: LSNC-construction} for the order ``$\prec$'')\footnote{Here, we can see that the reverse order of ``$\prec$'' is a topological order on the reversed graph $\mG^\top$ that is consistent with the natural partial order of the edges in $\mE$ on $\mG^\top$. The use of this reverse order is convenient for the subsequent discussion.} by
 \begin{align}\label{equ_f_e}
        f_e=\sum_{d\in \ein^{\mG^\top}\!(\tail^{\mG^\top}\!(e))}k_{d,e}\cdot f_d
 \end{align}
with the boundary condition that $f_{d_i'}$, $1 \leq i \leq R$ form the standard basis of the vector space $\Fq^R$.
\end{itemize}
%Then, the $\Fq$-valued rate-$R$ linear network code $\mbC_{\mN^\top}$ on $\mN^\top$ is written as
%\begin{align*}
%\mbC_{\mN^\top}=\big\{ K_v:\,v\in \mV\setminus S;~F_{\sigma_i}:\,\sigma_i\in S \big\}.
%\end{align*}
\end{defn}

We use $y_e$ to denote the symbol transmitted on $e$, $\forall~e\in \ein^{\mG^\top}\!(\rho)\bigcup \mE$. With $y_{d_i'}=x_i$ $(= \mathbf{x} \cdot f_{d'_i})$, $1\leq i \leq R$, each $y_e$ for $e\in \mE$ can be calculated recursively according to the reverse order of ``$\prec$'' on the edges in $\mE$ by
\begin{align*}%\label{equ_y_e}
y_e=\sum_{d\in \ein^{\mG^\top}\!(v)}k_{d,e}\cdot y_d,
\end{align*}
%or equivalently,
%\begin{align*}
%\big( y_e:~e\in \eout^{\mG^\top}\!(v) \big) = \big( y_d:~d\in \ein^{\mG^\top}\!(v) \big)\cdot K_v,
%\end{align*}
where $v=\tail^{\mG^\top}\!(e)$, and
together with \eqref{equ_f_e}, we have
\begin{align*}
y_e= \mathbf{x} \cdot f_{e}, \quad \forall\,e\in \mE.
\end{align*}
For each sink node $\sigma_i\in S$, we further have
\begin{align*}
\big( y_e:~e\in \ein^{\mG^\top}\!(\sigma_i) \big) = \mathbf{x} \cdot \left[ f_e:~e\in \ein^{\mG^\top}\!(\sigma_i) \right]
= \mathbf{x} \cdot F_{\sigma_i}.
\end{align*}
Then, the source message $\mathbf{x}$ can be decoded at $\sigma_i$ if and only if $\Rank\big(F_{\sigma_i}\big)=R$. Accordingly, the rate-$R$ code $\mbC_{\mN^\top}$ is called {\em decodable} if $\Rank\big(F_{\sigma_i}\big)=R$, $\forall~\sigma_i\in S$, or equivalently, for each decoding matrix $F_{\sigma_i}$,
there exists an $\Fq$-valued $\big\lvert\ein^{\mG^\top}\!(\sigma_i)\big\rvert \times R$ matrix $K_{\sigma_i}$ such that
\begin{align}\label{equ2-append}
F_{\sigma_i}\cdot K_{\sigma_i}=I_R,
\end{align}
where $I_R$ stands for the $R\times R$ identity matrix.

Furthermore, we set $k_{d, e}=0$ for all non-adjacent edge pairs $(d, e)$ on $\mG^\top$ which satisfy $\head^{\mG^\top}\!(d) \neq \tail^{\mG^\top}\!(e)$. We now define an $|\mE| \times |\mE|$ matrix
\begin{align*}
K = \big( k_{d, e} \big)_{d \in \mE,~e \in \mE}.
\end{align*}
We can see that according to the reverse order of ``$\prec$'', $K$ is lower triangular and all the diagonal elements are equal to zero. For an edge subset $\eta\subseteq \mE$ on $\mG^\top$, we let
\begin{align*}
\mathds{1}_{\eta} \triangleq \left[ \vec{1}_{e}:~e \in \eta,~~\bzero:~e\in \mE \setminus \eta \right]
\end{align*}
be an $ |\eta| \times |\mE|$ matrix in which the rows are indexed by the edges in $\eta$ and the columns are indexed by the edges in $\mE$, where $\vec{1}_{e}$ is the column $|\eta|$-vector whose component indexed by the edge $e$ is equal to~$1$ while all other components are equal to $0$, and $\bzero$ is an all-zero column $|\eta|$-vector. We readily see that the submatrix $\left[ \vec{1}_{e}:~e \in \eta \right]$ of $\mathds{1}_\eta$ is the $|\eta|\times |\eta|$ identity matrix. Koetter and M\'{e}dard \cite{Koetter-Medard-algebraic} proved the following equality:
\begin{align}\label{K-M-formula}
\Big[ f_e:~e \in \mE \Big]=K_{\rho}\cdot \mathds{1}_{\eout^{\mG^\top}\!(\rho)}\cdot\big(I-K\big)^{-1}.
\end{align}
Then, for each sink node $\sigma_i\in S$ on $\mG^{\top}$, by \eqref{K-M-formula}, we have
\begin{align}\label{equ3-append}
F_{\sigma_i}=\left[f_e:~e\in \ein^{\mG^\top}\!(\sigma_i)\right]&=\Big[ f_e:~e \in \mE \Big]\cdot \mathds{1}_{\ein^{\mG^\top}\!(\sigma_i)}^{\top}\\
&=K_{\rho}\cdot \mathds{1}_{\eout^{\mG^\top}\!(\rho)}\cdot\big(I-K\big)^{-1}\cdot\mathds{1}_{\ein^{\mG^\top}\!(\sigma_i)}^{\top}.
\end{align}
Together with~\eqref{equ2-append}, we obtain that
\begin{align}\label{equ4-append}
K_{\rho} \cdot \mathds{1}_{\eout^{\mG^\top}\!(\rho)} \cdot \big(I-K\big)^{-1} \cdot \mathds{1}_{\ein^{\mG^\top}\!(\sigma_i)}^{\top}\cdot K_{\sigma_i}=I_R,\quad \forall~\sigma_i\in S.
\end{align}

Next, we construct an $\Fq$-valued admissible $(R, 1)$ linear network code $\mbC$ for $(\mN, f)$ by ``reversing'' the decodable rate-$R$ linear network code $\mbC_{\mN^\top}$ on $\mN^\top$. Now, we consider the original graph $\mG$.
\begin{itemize}
  \item First, for each source node $\sigma_i\in S$, we let
\begin{align*}
  \begin{bmatrix}
    A_{i,e}:~e \in \eout(\sigma_i)
  \end{bmatrix}
  =K_{\sigma_i}^\top
\end{align*}
(cf.~\eqref{equ_R_1_lnc} for $A_{i,e}$). Recall from \eqref{equ1-append} that $\eout(\sigma_i)=\ein^{\mG^{\top}}\!(\sigma_i)$, which immediately implies that
\begin{align}\label{equ5-append}
A_{\sigma_i} = K_{\sigma_i}^\top \cdot \mathds{1}_{\ein^{\mG^\top}\!(\sigma_i)}
\end{align}
(cf.~\eqref{equ_A-sigma-i} for $A_{\sigma_i}$).
  \item For each intermediate node $v \in \mV \setminus (S \cup \{\rho\})$, we let $A_v=K_v^{\top}$, i.e., let
$a_{d,e}=k_{e,d}$ for all the adjacent pairs $(d,e)$ on $\mG$ with $\head(d)=\tail(e)=v$ (cf.~\eqref{equ_A-v} for $A_v$).
\end{itemize}

Now, by the above discussion, we have
\begin{align}\label{equ6-append}
A = K^\top
\end{align}
(cf.~\eqref{equ_A} for $A$). For the sink node $\rho$ on $\mG$, we consider
\begin{align}
G_{\rho}^{(\sigma_i)}&\triangleq
  \begin{bmatrix}
\vg_{e}^{\,(\sigma_i)}:~e\in \ein(\rho)
  \end{bmatrix}\nonumber\\
%= \begin{bmatrix}
%\vg_{e}^{\,(\sigma_i)}:~e\in \eout^{\mG^\top}\!(\rho)
%  \end{bmatrix}\nonumber\\
&=\begin{bmatrix}
    \vg_{e}^{\,(\sigma_i)}:~e\in \mE
  \end{bmatrix} \cdot \mathds{1}^\top_{\ein(\rho)}\nonumber\\
& = A_{\sigma_i} \cdot \big( I-A \big)^{-1} \cdot \mathds{1}^\top_{\ein(\rho)}\label{equ7-append}\\
& = K_{\sigma_i}^\top \cdot \mathds{1}_{\ein^{\mG^\top}\!(\sigma_i)} \cdot \big( I-K^\top \big)^{-1} \cdot \mathds{1}^\top_{\eout^{\mG^\top}\!(\rho)}\label{equ8-append}
\end{align}
(cf.~\eqref{ge-vec-form} for $\vg_{e}^{\,(\sigma_i)}$), where \eqref{equ7-append} follows from \eqref{G_rho_sigma}, and \eqref{equ8-append} follows from \eqref{equ5-append}, \eqref{equ6-append} and the equality $\mathds{1}^\top_{\eout^{\mG^\top}\!(\rho)}=\mathds{1}^\top_{\ein(\rho)}$ since $\eout^{\mG^\top}\!(\rho)=\ein(\rho)$.
Further, we obtain that
\begin{align*}
G_{\rho}^{(\sigma_i)} \cdot K_{\rho}^\top
& = K_{\sigma_i}^\top \cdot \mathds{1}_{\ein^{\mG^\top}\!(\sigma_i)} \cdot \big( I-K^\top \big)^{-1} \cdot \mathds{1}_{\eout^{\mG^\top}\!(\rho)}^\top \cdot K_{\rho}^\top \\
& = K_{\sigma_i}^\top \cdot \mathds{1}_{\ein^{\mG^\top}\!(\sigma_i)} \cdot \Big[ \big( I-K \big)^{-1}\Big]^\top \cdot \mathds{1}_{\eout^{\mG^\top}\!(\rho)}^\top \cdot K_{\rho}^\top \\
& = \left[ K_{\rho}\cdot \mathds{1}_{\eout^{\mG^\top}\!(\rho)}\cdot\big(I-K\big)^{-1}\cdot\mathds{1}_{\ein^{\mG^\top}\!(\sigma_i)}^{\top} \cdot K_{\sigma_i}\right]^\top= I_R,
\end{align*}
where the last equality follows from the equality \eqref{equ4-append}. Therefore, by considering all the source nodes $\sigma_i$ in $S$, we obtain that
\begin{align}\label{equ9-append}
G_{\rho} \cdot K_{\rho}^\top
%=
%  \begin{bmatrix}
%    \vg_{e}:~e\in \ein(\rho)
%  \end{bmatrix} \cdot K_{\rho}^\top
    =
  \begin{bmatrix}
    G_{\rho}^{\,(\sigma_1)} \\
    G_{\rho}^{\,(\sigma_2)} \\
    \vdots \\
    G_{\rho}^{\,(\sigma_s)}
  \end{bmatrix} \cdot K_{\rho}^\top
  =
   \begin{bmatrix}
    G_{\rho}^{\,(\sigma_1)}\cdot K_{\rho}^\top \\
    G_{\rho}^{\,(\sigma_2)}\cdot K_{\rho}^\top \\
    \vdots \\
    G_{\rho}^{\,(\sigma_s)}\cdot K_{\rho}^\top
  \end{bmatrix}
  =
   \begin{bmatrix}
  I_R \\
  I_R \\
  \vdots \\
  I_R
  \end{bmatrix}_{Rs\times R},
\end{align}
which, together with \eqref{equ_linear-global-encoding-vec}, implies that the sink node $\rho$ computes the algebraic sum with zero error $R$ times by using the code $\mbC$ once on the network $\mN$, i.e., $\mbC$ is an $\Fq$-valued admissible $(R,1)$  linear network code for the model $(\mN,f)$.

%%%%%%%%%%%%%%%%%%%%%%%%%%%%%%%%%%%%%%%%%%%%%%%%%

\end{document}